\documentclass[aps,prl,twocolumn,floatfix,nofootinbib,superscriptaddress,longbibliography,nobibnotes]{revtex4-2}

\usepackage{amsmath}
\usepackage{amssymb}
\usepackage{mathtools}

\usepackage[T1]{fontenc}
\usepackage{amsfonts}
\usepackage{newtxtext}
\usepackage[varvw]{newtxmath}
\usepackage{dsfont}                 
\usepackage{bbold}                  
\usepackage[normalem]{ulem}         

\usepackage{array}
\usepackage{dcolumn}                

\usepackage{enumerate}
\usepackage[shortlabels]{enumitem}

\usepackage[usenames,dvipsnames,table]{xcolor}
\usepackage{graphicx}
\graphicspath{{figs/}} 

\usepackage{physics}                
\usepackage{csquotes}               
\usepackage{multirow}               
\usepackage{booktabs}               

\usepackage[caption=false]{subfig}
\usepackage[colorlinks=true,citecolor=blue,linkcolor=blue,filecolor=blue]{hyperref}
\usepackage[capitalize]{cleveref}

\usepackage{orcidlink}

\newcommand{\sectitle}[1]{\emph{{#1}.---}}

\definecolor{JM}{rgb}{1,0.25,0.9}

\renewcommand{\c}[1]{\mathcal{#1}}
\newcommand{\vol}{\mathop{\mathrm{vol}}}
\newcommand{\BZ}{{\mathrm{BZ}_N}}
\newcommand{\YM}{{\textrm{YM}}}

\newcommand{\nn}{\nonumber}

\renewcommand{\geq}{\geqslant}
\renewcommand{\leq}{\leqslant}
\renewcommand{\hom}{{\mathrm{Hom}}}

\begin{document}

\title{Hyperbolic lattices and two-dimensional Yang-Mills theory}

\author{G. Shankar}
\email{sankaran@ualberta.ca}
\affiliation{Department of Physics, University of Alberta, Edmonton, Alberta T6G 2E1, Canada}

\author{Joseph Maciejko\,\orcidlink{0000-0002-6946-1492}}
\email{maciejko@ualberta.ca}
\affiliation{Department of Physics, University of Alberta, Edmonton, Alberta T6G 2E1, Canada}
\affiliation{Theoretical Physics Institute \& Quantum Horizons Alberta, University of Alberta, Edmonton, Alberta T6G 2E1, Canada}

\date{\today}

\begin{abstract}
Hyperbolic lattices are a new type of synthetic quantum matter emulated in circuit quantum electrodynamics and electric-circuit networks, where particles coherently hop on a discrete tessellation of two-dimensional negatively curved space. While real-space methods and a reciprocal-space hyperbolic band theory have been recently proposed to analyze the energy spectra of those systems, discrepancies between the two sets of approaches remain. In this work, we reconcile those approaches by first establishing an equivalence between hyperbolic band theory and $U(N)$ topological Yang-Mills theory on higher-genus Riemann surfaces. We then show that moments of the density of states of hyperbolic tight-binding models correspond to expectation values of Wilson loops in the quantum gauge theory and become exact in the large-$N$ limit.
\end{abstract}

\maketitle

\sectitle{Introduction}Hyperbolic lattices represent a new class of synthetic quantum systems whereby local degrees of freedom live on a discrete tessellation of the hyperbolic plane, the two-dimensional (2D) hyperbolic space with constant negative curvature. The geometry of those lattices is encoded in the Schl\"afli symbol $\{p,q\}$ that denotes a tiling by regular $p$-gons, $q$ of which meet at each vertex of the lattice, where $(p-2)(q-2)>4$ requires negative curvature~\cite{boettcher2022}. Experimental realizations include the circuit-quantum-electrodynamics implementation of the kagome-like line graph of the $\{7,3\}$ lattice~\cite{kollar2019}, and classical electric-circuit simulations of quantum hopping on the $\{3,7\}$~\cite{lenggenhager2021}, $\{6,4\}$~\cite{zhang2022,pei2023}, and $\{8,8\}$~\cite{zhang2023} lattices. Besides garnering interest as a novel type of condensed-matter system~\cite{zhang2022,pei2023,zhang2023,Daniska:2016,Yu:2020,Zhu:2021,Liu:2022,urwyler2022,chen2023,Tao:2022,Liu:2023,bzdusek2022,Mosseri:2022,Stegmaier:2021,Bienias:2022,chen2023a,Gluscevich:2023,gluscevich2023b,gotz2024,lenggenhager2024,dusel2024}, hyperbolic lattices have also been explored recently as quantum error-correcting codes~\cite{pastawski2015,breuckmann2016,breuckmann2017,lavasani2019,jahn2021} and platforms for simulating the holographic principle~\cite{boettcher2020,boyle2020,asaduzzaman2020,brower2021,basteiro2022,basteiro2023,basteiro2022b,Chen:2023c,dey2024}.

Although a single-particle problem, computing the energy spectrum of a tight-binding model on a hyperbolic lattice is nontrivial on account of the curvature of space, which precludes a naive application of the well-known Bloch theorem~\cite{ashcroft1976}. Hyperbolic band theory (HBT)~\cite{maciejko2021,maciejko2022} proposes a generalization of this theorem whereby eigenstates acquire a Bloch factor valued in the unitary group $U(N)$ under non-Euclidean lattice translations. Those translations form the infinite \emph{Fuchsian group} $\Gamma$, which plays a role akin to the translation group of ordinary lattices but is noncommutative. The matrix-valued phases correspond to $N$-dimensional unitary irreducible representations (irreps) of $\Gamma$ with $N\geq 1$. In the simplest case of $N=1$, referred to below as $U(1)$ HBT, the corresponding Abelian Bloch states live in a $2g$-dimensional Brillouin zone (BZ) torus, where $g$ is the genus of the compactified unit cell~\cite{maciejko2021}. Numerical diagonalization on finite lattices reveals good approximate agreement between real-space spectra and the predictions of $U(1)$ HBT for several hyperbolic tight-binding models~\cite{bzdusek2022,urwyler2022,chen2023a,chen2023}. However, for very large lattices, as well as for certain lattice models, the predictions of $U(1)$ HBT are qualitatively at odds with those of real-space methods, which include diagonalization on large periodic clusters~\cite{lux2022,lux2023,tummuru2023} and continued fraction expansions~\cite{mosseri2023}. Thus, \emph{non-Abelian} Bloch states with $N\geq 2$ are expected to play an important role in capturing the correct spectrum in the thermodynamic limit. The corresponding non-Abelian BZ is a moduli space of rank-$N$ holomorphic vector bundles~\cite{maciejko2022,kienzle2022}; by contrast with the Abelian BZ, it is a manifold of dimension $2N^2(g-1)+2$ with a non-toroidal geometry. Although limited classes of non-Abelian Bloch states may be accessed via periodic clusters~\cite{maciejko2022}, incomplete parameterizations~\cite{cheng2022}, random sampling~\cite{tummuru2023}, and induced representations~\cite{lenggenhager2023}, the lack of an explicit, complete parameterization of the non-Abelian BZ has precluded precise tests of the validity of HBT.

In this work, we show that $U(N)$ HBT captures the exact infinite-lattice spectrum in the $N\rightarrow\infty$ limit, implying that $N=1,2,3,\ldots$ can be viewed as a sequence of approximations which converges to the thermodynamic limit. To establish this conclusion, we first show that $U(N)$ HBT maps onto a quantum gauge field theory: two-dimensional Yang-Mills theory (2D YM) with gauge group $U(N)$, defined on a Riemann surface of genus $g\geq 2$ and in the limit of vanishing gauge coupling $e^2\rightarrow 0$~\cite{migdal1975,witten1991}. While the connection between HBT and the classical YM equations on a Riemann surface~\cite{atiyah1983} has been noticed before~\cite{maciejko2022,kienzle2022}, here we show that the full quantum YM gauge theory with dynamical $U(N)$ gauge field---i.e., pure 2D quantum chromodynamics (QCD$_2$), without matter---is key to establishing the exactitude of HBT in the large-$N$ limit. Integrals over the non-Abelian BZ become functional integrals over gauge field configurations, which can be computed using the exact solubility of 2D YM and its status as a topological quantum field theory (TQFT) in the $e^2\rightarrow 0$ limit~\cite{cordes1995}. Circumventing the BZ parameterization problem in this way, we finally show that moments of the density of states (DOS) of an arbitrary hopping model on a hyperbolic lattice can be expressed in terms of expectation values of Wilson loops in 2D YM, and converge to the exact moments of the infinite lattice in the $N\rightarrow\infty$ limit.

\begin{figure}[t!]
\centering
    \includegraphics[width=\columnwidth]{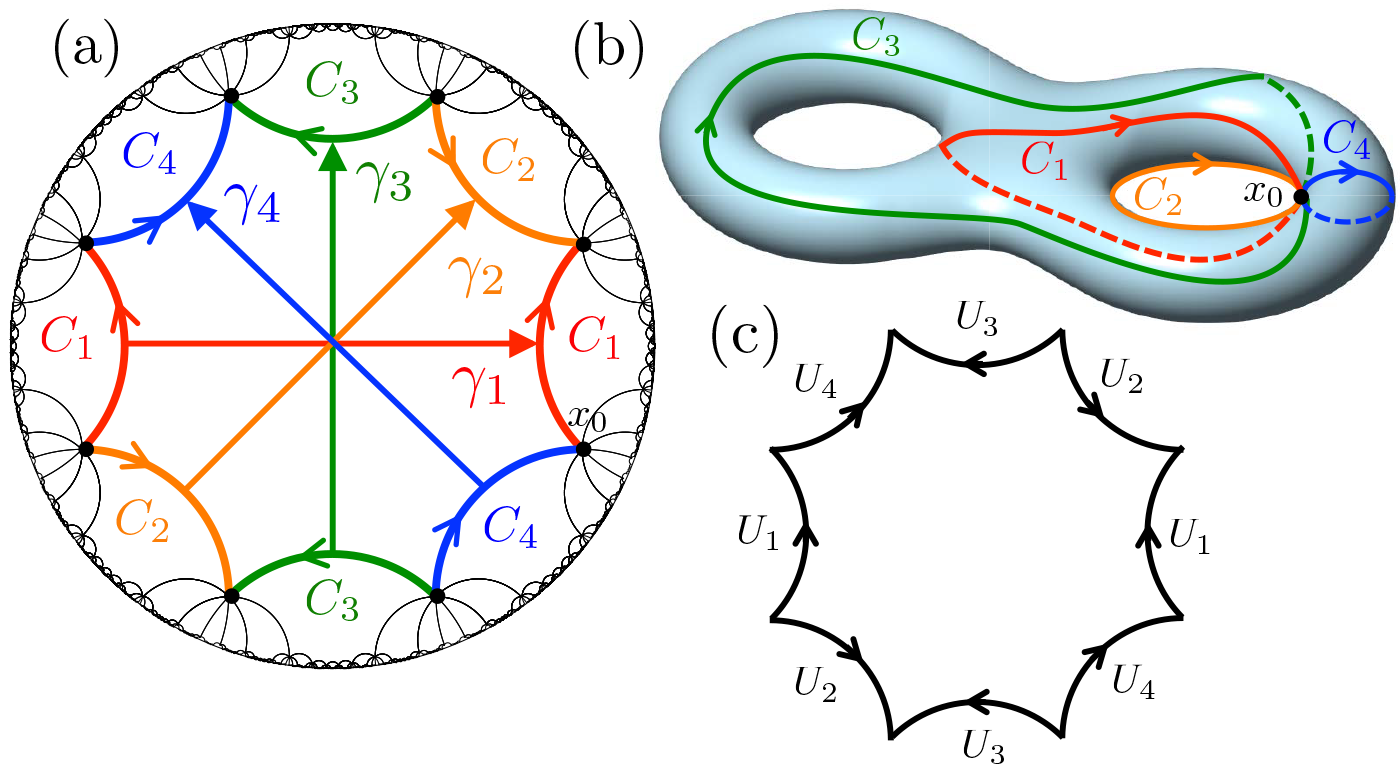}
    \caption[]{(a) The $\{8,8\}$ lattice, with the four generators $\gamma_1,\ldots,\gamma_4$ of its translation group $\Gamma$; (b) the compactified unit cell $\Sigma$, obtained by gluing pairwise the boundary segments $C_1,\ldots,C_4$ of the octagonal unit cell in (a). Under the isomorphism $\Gamma\cong\pi_1(\Sigma)$, each generator $\gamma_j$ corresponds to a noncontractible loop $C_j$ on $\Sigma$ based at $x_0$. (c) Migdal's lattice gauge theory, where $U(N)$ gauge variables $U_\ell$ live on the links $\ell$ of a discretization of $\Sigma$.}
\label{fig:bolza}
\end{figure}

\sectitle{Hyperbolic band theory}For simplicity, we focus in the main text on the nearest-neighbor hopping model on the $\{8,8\}$ lattice~\cite{maciejko2022,lux2022,mosseri2023,lenggenhager2023} which constitutes an illustrative example [Fig.~\ref{fig:bolza}(a)], but also explain how to generalize our ideas to arbitrary finite-range hopping models on any $\{p,q\}$ lattice~\cite{SM}. Translations from a reference site of the $\{8,8\}$ lattice to any other site form the infinite Fuchsian group $\Gamma$ of the lattice~\cite{Katok}; this gives a natural discrete position basis $\{\ket{\gamma},\gamma\in\Gamma\}$ labeled by group elements, which is complete and orthonormal. We are interested in the spectrum of the translationally invariant nearest-neighbor hopping Hamiltonian,
\begin{align}\label{H}
    \hat{H}=\sum_{\gamma\in\Gamma}\sum_\alpha\ket{\gamma\gamma_\alpha}\bra{\gamma},
\end{align}
where the set $\{\gamma_\alpha\}$ includes the four generators $\gamma_j$, $j=1,\ldots,4$ of $\Gamma$ and their inverses, which correspond to translations from the reference site to its 8 nearest neighbors~\cite{maciejko2021}. Explicitly, $\{\gamma_\alpha\}=\{\gamma_1,\gamma_2,\gamma_3,\gamma_4,\gamma_1^{-1},\gamma_2^{-1},\gamma_3^{-1},\gamma_4^{-1}\}$. The generators obey the single relation $\c{R}(\{\gamma_j\})=1$, where
\begin{align}\label{relator}
    \c{R}(\{\gamma_j\})=\gamma_1\gamma_2^{-1}\gamma_3\gamma_4^{-1}\gamma_1^{-1}\gamma_2\gamma_3^{-1}\gamma_4.
\end{align}

HBT corresponds to block-diagonalizing (\ref{H}) by introducing ``momentum'' or representation-basis kets $\{\ket{K,\lambda,\nu}\}$ whose position-space wave functions,
\begin{align}\label{Klambdanukets}
    \braket{\gamma}{K,\lambda,\nu}=D_{\nu\lambda}^{(K)}(\gamma)^*,\hspace{5mm}
    \lambda,\nu=1,\ldots,N,
\end{align}
are matrix elements of $N\times N$ irrep matrices $D^{(K)}(\gamma)$ for the group $\Gamma$~\cite{rotor}. Since $\Gamma$ is infinite, for any fixed $N=1,2,\ldots$, there is a continuous set $K\in\BZ$ of irreps of $\Gamma$ which forms the $U(N)$ {\it moduli space} (or character variety) $\BZ$~\cite{narasimhan1965}, interpreted in HBT as a generalized BZ~\cite{maciejko2022,kienzle2022}. Since any element of $\Gamma$ can be expressed in terms of the generators $\gamma_j$, $\BZ$ can be defined explicitly as the set of four $U(N)$ matrices $U_j\equiv D^{(K)}(\gamma_j)$, $j=1,\ldots,4$ that obey the group relation (\ref{relator}):
\begin{align}\label{BZdef}
    \BZ\equiv\{U_1,U_2,U_3,U_4\in U(N):\c{R}(\{U_j\})=1\},
\end{align}
where $\c{R}(\{U_j\})=U_1U_2^{-1}U_3U_4^{-1}U_1^{-1}U_2U_3^{-1}U_4$.

For each $N$, the basis $\{\ket{K,\lambda,\nu}\}$ block-diagonalizes the Hamiltonian (see Appendix A):
\begin{align}\label{HBlockDiag}
    \langle K,\lambda,\nu|\hat{H}|K',\lambda',\nu'\rangle=\delta(K-K')\delta_{\nu\nu'}\c{H}_{\lambda'\lambda}(K),
\end{align}
where
\begin{align}\label{HBloch}
    \c{H}(K)\equiv\sum_\alpha D^{(K)}(\gamma_\alpha),
\end{align}
is the $U(N)$ {\it Bloch Hamiltonian}, a $K$-dependent $N\times N$ Hermitian matrix. From this Hamiltonian, we define a $U(N)$ DOS,
\begin{align}\label{UNDOS}
    \rho_N(E)\equiv\frac{1}{N}\int_\BZ\frac{dK}{\vol(\BZ)}\tr\delta(E-\c{H}(K)),
\end{align}
which is normalized to unity, $\int dE\,\rho_N(E)=1$, defining $\vol(\BZ)\equiv\int_\BZ dK$ as the volume of the non-Abelian BZ. Here, integral over $K$ is defined explicitly as the $U(N)$ Haar integral over the four matrices $U_j$ obeying the nonlinear constraint (\ref{BZdef}). Equivalently, we define the $U(N)$ DOS moments
\begin{align}\label{moments}
    \rho_N^{(k)}\equiv\int dE\,\rho_N(E)E^k,\hspace{5mm}k=1,2,3,\ldots,
\end{align}
from which $\rho_N(E)$ can in principle be reconstructed~\cite{trias1983}.

To compute key physical observables like the DOS (\ref{UNDOS}) or its moments (\ref{moments}) in HBT, one must in principle be able to find explicit coordinates for the space $\BZ$ in Eq.~(\ref{BZdef}), evaluate the Bloch Hamiltonian (\ref{HBloch}) in those coordinates, and integrate over them. Those appear to be hopeless tasks: except for $N=1$, where $\BZ$ is a simple 4D torus~\cite{maciejko2021}, those spaces have a complicated geometry for which one does not expect to find analytically tractable, global coordinates~\cite{thaddeus1995}. Brute-force approaches like random numerical sampling of $\BZ$~\cite{tummuru2023} obviate the need for coordinates but face uncertainties regarding whether the sampling is appropriately uniform (i.e., with the right choice of measure).

\sectitle{Topological Yang-Mills theory}We now show that the issues above can be resolved by mapping HBT to an exactly soluble quantum gauge theory, 2D YM~\cite{migdal1975,witten1991}. First, we show that $\vol(\BZ)$ is given by the partition function of 2D YM. From Eq.~(\ref{BZdef}), we have
\begin{align}
    \vol(\BZ)=\prod_{j=1}^4\int dU_j\,\delta\bigl(\c{R}(\{U_j\}),1\bigr),
\end{align}
where $\delta(U,V)$ is a delta function on the $U(N)$ group manifold. Using the Peter-Weyl theorem (see Appendix B), this delta function can be expressed as a sum over all irreps $R$ of $U(N)$, with dimensions $d_R$ and characters $\chi_R(U)$:
\begin{align}\label{deltaUN}
\delta(U,V)=\sum_R d_R\chi_R(UV^{-1}),    
\end{align}
hence
\begin{align}\label{volBZN}
   \vol(\BZ)=\prod_{j=1}^4\int dU_j\sum_Rd_R\chi_R\bigl(\c{R}(\{U_j\})\bigr). 
\end{align}

Originally derived by Migdal~\cite{migdal1975}, the partition function of 2D YM on a Euclidean spacetime lattice with oriented links $\ell$ and plaquettes $P$ is $Z_\YM(e^2)=\prod_\ell\int dU_l\prod_P W_P$. Here, $e^2$ is the gauge coupling, $U_\ell$ is a $U(N)$ group element living on link $\ell$, and $W_P$ is the plaquette Boltzmann weight, given by
\begin{align}
    W_P=\sum_Rd_R\chi_R\bigl(\textstyle\prod_{\ell\in P}U_\ell\bigr)e^{-e^2 A_Pc_2(R)/2},
\end{align}
where $A_P$ is the area of plaquette $P$ and $c_2(R)$ is a representation-dependent constant. Comparing with Eq.~(\ref{volBZN}), we obtain our first main result: $\vol(\BZ)=Z_\YM(0)$. The volume of the non-Abelian BZ is given by the partition function of $U(N)$ 2D YM in the limit of vanishing gauge coupling $e^2=0$, computed on the genus-2 Riemann surface $\Sigma$ corresponding to the compactified unit cell of the $\{8,8\}$ lattice [Fig.~\ref{fig:bolza}(b)]. Indeed, the spacetime lattice here contains a single octagonal plaquette $P$ with 8 pairwise identified links and link variables $U_1,\ldots,U_4$ [Fig.~\ref{fig:bolza}(c)], allowing the identification $\c{R}(\{U_j\})=\prod_{\ell\in P}U_\ell$.

When $e^2=0$, 2D YM is in fact a TQFT~\cite{birmingham1991}, which can be seen in two ways. First, the Haar integrals over $U_j$ in Eq.~(\ref{volBZN}) can be performed explicitly using character identities~\cite{SM}; this gives
\begin{align}\label{volBZ}
\vol(\BZ)=Z_\YM(0)=\sum_R d_R^{\chi(\Sigma)},
\end{align}
where $\chi(\Sigma)=-2$ is the Euler characteristic of $\Sigma$. Thus, the partition function only depends on the topology of $\Sigma$. Second, Migdal's lattice gauge theory (for any $e^2$) exhibits the property of \emph{subdvision invariance}, i.e., the partition function is independent of the level of discretization of spacetime~\cite{SM}. Thus, Migdal's partition function is equal (with a suitable choice of gauge-fixing procedure) to that of \emph{continuum} 2D YM~\cite{witten1991}:
\begin{align}\label{ZYM}
    Z_\YM(e^2)=\int\c{D}A\,e^{-\frac{1}{2e^2}\int_\Sigma\tr F\wedge \star F},
\end{align}
where $A$ is the gauge field 1-form valued in the Lie algebra $\mathfrak{u}(N)$, $F=\dd A+A\wedge A$ is the field strength 2-form, and $\star$ is the Hodge star. Introducing a $\mathfrak{u}(N)$-valued Hubbard-Stratonovich 0-form field $\phi$, Eq.~(\ref{ZYM}) can be written as
\begin{align}\label{ZYM2}
    Z_\YM(e^2)=\int\c{D}A\c{D}\phi\,e^{-\frac{e^2}{2}\int_\Sigma\omega\tr\phi^2-i\int_\Sigma\tr\phi F},
\end{align}
where $\omega$ is a volume form on $\Sigma$. In the limit $e^2\rightarrow 0$, Eq.~(\ref{ZYM2}) does not depend on the choice of volume form or metric on $\Sigma$, and is thus topological.

Our mapping to topological 2D YM additionally reveals that the non-Abelian BZ can be viewed as a moduli space of flat $U(N)$ connections on $\Sigma$, i.e., those with $F=0$~\cite{narasimhan1965,maciejko2022,kienzle2022}. Indeed, in the limit $e^2\rightarrow 0$, integrating out $\phi$ in Eq.~(\ref{ZYM2})
produces a functional delta function which localizes the integral over all $A$ to the subspace with $F=0$:
\begin{align}\label{ZYMtop}
    Z_\YM(0)=\int\c{D}A\,\delta[F].
\end{align}
Up to an overall proportionality constant, this volume corresponds to a canonical choice of measure on $\BZ$, the co-called Atiyah-Bott-Goldman measure~\cite{witten1991,atiyah1983,goldman1984}. In the language of connections, the matrices $U_j$ that define $\BZ$ in Eq.~(\ref{BZdef}) correspond to nontrivial holonomies of $A$ along noncontractible loops $C_j$ on $\Sigma$ [Fig.~\ref{fig:bolza}(b)]:
\begin{align}\label{holonomy}
    U_j=D^{(K)}(\gamma_j)=\c{P}e^{i\oint_{C_j}A},
\end{align}
where $\c{P}$ denotes path ordering. (For a flat connection, holonomies along contractible cycles are trivial.) Mathematically, this identification is known as the Riemann-Hilbert correspondence~\cite{narasimhan1965,garcia-raboso2015,kienzle2022}, i.e., the isomorphism $\Gamma\cong\pi_1(\Sigma)$ between the Fuchsian group $\Gamma\ni\gamma_j$ of the lattice and the fundamental group $\pi_1(\Sigma)\ni C_j$ of the Riemann surface which ensures the holonomies (\ref{holonomy}) obey the group relation $\c{R}(\{U_j\})=1$. Physically, the geometry of the infinite lattice can be encoded in that of its \emph{compactified unit cell} $\Sigma$. For $N=1$, path ordering is unnecessary and we recover $D^{(K)}(\gamma_j)=e^{ik_j}$ with the $2\pi$-periodic hyperbolic crystal momenta $k_j=\oint_{C_j}A$~\cite{maciejko2021}; we henceforth focus on $N\geq 2$. We stress that, while connections and their holonomies can be obtained by solving the classical YM equations on $\Sigma$~\cite{atiyah1983}, integrating over all possible (flat) connections $K\in\BZ$---which is necessary to calculate physical observables such as the DOS, Eq.~(\ref{UNDOS})---amounts to performing a functional integral in \emph{quantum} 2D YM, i.e., QCD$_2$.

\sectitle{DOS moments from Wilson loops}Having shown that the volume of the non-Abelian BZ can be computed as the partition function of topological 2D YM on a higher-genus Riemann surface, we next show that the $U(N)$ DOS moments (\ref{moments}) correspond to expectation values of Wilson loop operators. In Appendix A, we express the $k$th moment as
\begin{align}\label{kthmoment}
    \rho_N^{(k)}=\frac{1}{N}\sum_{\alpha_1}\cdots\sum_{\alpha_k}\int_\BZ\frac{dK}{\vol(\BZ)}\tr D^{(K)}(\gamma_{\alpha_1}\cdots\gamma_{\alpha_k}).
\end{align}
Comparing with Eq.~(\ref{holonomy}), we see appearing the holonomy along a composite loop $C\equiv C_{\alpha_1}\cdots C_{\alpha_k}$ to which the translation $\gamma\equiv\gamma_{\alpha_1}\cdots\gamma_{\alpha_k}$ maps under the isomorphism $\Gamma\cong\pi_1(\Sigma)$. Since $D^{(K)}$ is a $N$-dimensional unitary irrep, in the 2D YM interpretation, the trace of this holonomy appearing in Eq.~(\ref{kthmoment}) is nothing but the gauge-invariant Wilson loop operator $W(C)$ in the fundamental representation of $U(N)$:
\begin{align}
    W(C)\equiv\tr_F\bigl(\c{P}e^{i\oint_C A}\bigr)=\tr D^{(K)}(\gamma).
\end{align}
Since $\vol(\BZ)=Z_\YM(0)$, integration over the non-Abelian BZ normalized by the BZ volume corresponds to a quantum expectation value of $W(C)$ in topological 2D YM:
\begin{align}\label{moment1}
    \rho_N^{(k)}=\frac{1}{N}\sum_{\alpha_1}\cdots\sum_{\alpha_k}\langle W(C_{\alpha_1}\cdots C_{\alpha_k})\rangle.
\end{align}
Crucially, since we are computing expectation values in a TQFT, $\langle W(C)\rangle$ only depends on the homotopy class of $C$.

The computation of Wilson-loop expectation values in 2D YM is a rich subject in mathematical physics~\cite{witten1991,cordes1995,makeenko1979,kazakov1980,kazakov1981,rusakov1990,dahlqvist2022,blau1992,thompson1993,magee2022,magee2021}. To compute Eq.~(\ref{moment1}), we divide the loops $C=C_{\alpha_1}\cdots C_{\alpha_k}$ appearing in the sum into three categories with respect to homotopy: contractible loops, homologically nontrivial loops, and homologically trivial noncontractible loops~\cite{blau1992,thompson1993}. Contractible loops can be contracted to a point. They are in the trivial homotopy class, for which the holonomy is the identity element ($\gamma=1$) and its trace in the fundamental representation gives $\langle W(C)\rangle=N$. Homologically nontrivial loops are those for which $\gamma\notin[\Gamma,\Gamma]$. Here, $[\Gamma,\Gamma]$ is the commutator subgroup---the set of elements of $\Gamma$ that would reduce to the identity if the noncommutativity of the group were ignored. In Appendix C, we show that $\langle W(C)\rangle$ vanishes identically for such loops, for all $N$.

\sectitle{The large-$N$ limit}Finally, we turn to homologically trivial noncontractible loops, for which $\gamma\in[\Gamma,\Gamma]$ but $\gamma\neq 1$. For such loops, the Wilson-loop expectation value behaves as $\langle W(C)\rangle\sim\c{O}(1)$ in the large-$N$ limit (see Appendix C). Since as shown above, $\langle W(C)\rangle$ vanishes identically for $\gamma\notin[\Gamma,\Gamma]$, and $\langle W(C)\rangle=N$ for $\gamma=1$, the $k$th $U(N)$ DOS moment (\ref{moment1}) becomes
\begin{align}\label{moment-largeN}
    \rho_N^{(k)}=\rho_\infty^{(k)}+\c{O}(1/N),
\end{align}
where the $\c{O}(1/N)$ contribution is from homologically trivial noncontractible loops. The $\c{O}(1/N^0)$ contribution, which arises from contractible Wilson loops, is
\begin{align}\label{moment-infinity}
    \rho_\infty^{(k)}=\sum_{\alpha_1}\cdots\sum_{\alpha_k}\delta_{\gamma_{\alpha_1}\cdots\gamma_{\alpha_k},1},
\end{align}
i.e., the number of words of length $k$ in the generators $\gamma_j\in\Gamma$ that are equal to the identity element. On the one hand, this algebraic word problem~\cite{Robinson} has a geometric interpretation: a word of length $k$ that is equal to the identity corresponds to a closed loop, or cycle, of length $k$ on the Cayley graph of $\Gamma$. On the other hand, the $k$th DOS moment for hopping on a graph corresponds precisely to the average number of cycles of length $k$ on this graph. Since the $\{8,8\}$ hyperbolic lattice is nothing but the Cayley graph of $\Gamma$, we conclude that the $k$th moment of the {\it exact} thermodynamic-limit DOS $\rho(E)$ for the nearest-neighbor hopping Hamiltonian (\ref{H}) is given by the large-$N$ limit of the $k$th moment (\ref{moment-largeN}) computed in $U(N)$ HBT:
\begin{align}
    \rho^{(k)}\equiv\int dE\,\rho(E)E^k=\lim_{N\rightarrow\infty}\rho_N^{(k)},
\end{align}
where the limit is taken for fixed $k$. Although we have arrived at this result here for the $\{8,8\}$ lattice, our conclusion holds more generally~\cite{SM}. An arbitrary $\{p,q\}$ lattice can be described as a hyperbolic Bravais lattice with a finite sublattice basis~\cite{boettcher2022,chen2023}. The compactified Bravais unit cell $\Sigma$ is a compact Riemann surface whose genus $g\geq 2$ depends on $p$ and $q$. The volume of $\BZ$ is again Eq.~(\ref{volBZ}) but with $\chi(\Sigma)=2-2g$. The $k$th $U(N)$ DOS moment is a sum of terms that factorize into a purely geometric piece, the Wilson-loop expectation value $\langle W(C_{\alpha_1}\cdots C_{\alpha_k})\rangle$, and a piece $\tr T(\gamma_{\alpha_1})\cdots T(\gamma_{\alpha_k})$, where $T(\gamma_\alpha)$ is a finite-dimensional hopping matrix in the space of sublattice, orbital, and/or spin degrees of freedom. Here, $\alpha$ indexes a finite set of $\Gamma$-translations that, physically, can capture hopping terms of arbitrary range. In this set, each element appears together with its inverse, and is expressible in terms of the identity and the $2g$ generators of $\Gamma$. The large-$N$ limit applies only to the geometric piece and again yields the exact thermodynamic-limit DOS moments.

\sectitle{Conclusion}In summary, we have shown that $U(N)$ HBT captures the exact infinite-lattice density of states in the limit $N\rightarrow\infty$. To establish this result, we mapped this theory onto an exactly soluble quantum gauge theory, topological 2D YM (pure QCD$_2$) with gauge group $U(N)$. Observables in HBT map onto those of the gauge theory: the volume of the non-Abelian Brillouin zone is the 2D YM partition function, and DOS moments are related to Wilson-loop expectation values. In the large-$N$ limit, only contractible Wilson loops contribute to the DOS moments, and the latter recover their exact value describing the geometry of the infinite hyperbolic lattice. Our results suggest that, as applied to the thermodynamic limit, $U(N)$ HBT for $N=1,2,3,\ldots$ is better viewed as a converging sequence of approximations than as a way to capture distinct superselection sectors of the infinite-lattice Hilbert space. This point of view accords with, and gives a physical interpretation of, the $C^*$-algebraic perspective on the infinite-lattice $\ell^2$-spectrum as obtained from a converging sequence of finite-dimensional representations of $\Gamma$~\cite{lux2022}. The BZ of an infinite hyperbolic lattice is thus formally a moduli space $\mathrm{BZ}_\infty$ of infinite-dimensional irreps which, despite being itself infinite dimensional, has finite volume: Eq.~(\ref{volBZ}) gives $\vol(\mathrm{BZ}_\infty)=1$ since only the trivial irrep with $d_R=1$ contributes in the large-$N$ limit~\cite{rusakov1993}. Our work is also consistent with other recent mathematical work~\cite{nagy2022}, according to which $U(N)$ HBT in the $N\rightarrow\infty$ limit has the properties expected of a well-behaved Bloch transform, i.e., injectivity and (asymptotic) unitarity.

In addition to resolving the apparent discrepancies between real-space approaches to hyperbolic lattices and HBT, our work reveals an unexpected link between the physics of synthetic condensed matter and topological gauge theories studied in high-energy physics. The connection between QCD$_2$ and string theory~\cite{gross1993,*gross1993b,cordes1997} is worthy of reexamination in the current context, and may reveal interesting stringy interpretations of hyperbolic-lattice physics, or suggest experimental table-top simulations of stringy phenomena. On the theoretical side, large-$N$ techniques inspired by conformal field theory, developed originally in the 2D YM context~\cite{douglas1993}, might also be profitably applied to the computation of physical properties of hyperbolic lattices. Finally, having identified $\mathrm{BZ}_\infty$ as the relevant reciprocal space for infinite hyperbolic lattices, it would be interesting to compute topological invariants~\cite{lux2023,sun2024} of hyperbolic bandstructures as invariants of Bloch Hamiltonians $\c{H}(K)$ over $\mathrm{BZ}_\infty$, as done in ordinary topological band theory~\cite{chiu2016}, and eventually develop a complete classification of topological phases of hyperbolic quantum matter.

\bigskip

\sectitle{Acknowledgements}We thank I.~Boettcher, T.~Bzdu\v{s}ek, A.~Chen, T.~Creutzig, S.~Dey, P.~Lenggenhager, T.~Neupert, E.~Prodan, S.~Rayan, and C.~Sun for useful discussions. G.~S. was supported by the Golden Bell Jar Scholarship in Physics and the Alberta Graduate Excellence Scholarship. J.~M.~was supported by NSERC Discovery Grants \#RGPIN-2020-06999 and \#RGPAS-2020-00064; the Canada Research Chair (CRC) Program; the NSERC Alliance-Alberta Innovates Advance Program; and the Pacific Institute for the Mathematical Sciences (PIMS) Collaborative Research Group program.

\bigskip

\setcounter{equation}{0}
\renewcommand{\theequation}{A\arabic{equation}}

\sectitle{Appendix A: Hyperbolic band theory}Equation~(\ref{HBlockDiag}) follows from the definitions (\ref{Klambdanukets}), (\ref{HBloch}) and the representation property $D^{(K)}(\gamma\gamma')=D^{(K)}(\gamma)D^{(K)}(\gamma')$:
\begin{align}
        &\langle K,\lambda,\nu|\hat{H}|K',\lambda',\nu'\rangle=\sum_{\gamma\in\Gamma}\sum_{\alpha}\langle K,\lambda,\nu|\gamma\gamma_\alpha\rangle\langle\gamma|K',\lambda',\nu'\rangle\nn\\
&\hspace{20mm}=\sum_{\gamma\in\Gamma}\sum_{\alpha} D_{\nu\lambda}^{(K)}(\gamma\gamma_\alpha)D_{\nu'\lambda'}^{(K')}(\gamma)^*\nn\\
&\hspace{20mm}=\sum_{\gamma\in\Gamma}\sum_{\alpha}\sum_\mu D_{\nu\mu}^{(K)}(\gamma)D_{\mu\lambda}^{(K)}(\gamma_\alpha)D_{\nu'\lambda'}^{(K')}(\gamma)^*\nn\\
&\hspace{20mm}=\frac{|\Gamma|}{N}\delta_{KK'}\delta_{\nu\nu'}\c{H}_{\lambda'\lambda}(K),
\end{align}
using also the Schur orthogonality relation
\begin{align}
    \frac{N}{|\Gamma|}\sum_{\gamma\in\Gamma}D_{\nu\mu}^{(K)}(\gamma)D_{\nu'\lambda'}^{(K')}(\gamma)^*=\delta_{KK'}\delta_{\nu\nu'}\delta_{\mu\lambda'}.
\end{align}
Here, we have approximated $\Gamma$ as an arbitrarily large but finite group of order $|\Gamma|\gg 1$, corresponding to a finite lattice with periodic boundary conditions~\cite{maciejko2022}, such that $K,K'$ are discrete. In the thermodynamic limit $|\Gamma|\rightarrow\infty$, $K,K'\in\BZ$ become continuous variables and we define a delta function $\delta$ on $\BZ$ such that $\lim_{|\Gamma|\rightarrow\infty}(|\Gamma|/N)\delta_{KK'}=\delta(K-K')$, yielding Eq.~(\ref{HBlockDiag}).

From Eqs.~(\ref{HBloch}--\ref{moments}), Eqs.~(\ref{kthmoment}) and (\ref{moment1}) for the $k$th $U(N)$ DOS moment are derived as
\begin{align}
    &\rho_N^{(k)}=\int dE\,\rho_N(E)E^k\nn\\
    &=\frac{1}{N}\int_\BZ\frac{dK}{\vol(\BZ)}\int dE\,E^k\tr\delta(E-\c{H}(K))\nn\\
    &=\frac{1}{N}\int_\BZ\frac{dK}{\vol(\BZ)}\tr\c{H}(K)^k\nn\\
    &=\frac{1}{N}\int_\BZ\frac{dK}{\vol(\BZ)}\sum_{\alpha_1}\cdots\sum_{\alpha_k}\tr D^{(K)}(\gamma_{\alpha_1})\cdots D^{(K)}(\gamma_{\alpha_k})\nn\\
    &=\frac{1}{N}\int_\BZ\frac{dK}{\vol(\BZ)}\sum_{\alpha_1}\cdots\sum_{\alpha_k}\tr D^{(K)}(\gamma_{\alpha_1}\cdots\gamma_{\alpha_k})\nn\\
    &=\frac{1}{N}\int\frac{\c{D}A\,\delta[F]}{Z_\YM(0)}\sum_{\alpha_1}\cdots\sum_{\alpha_k}W(C_{\alpha_1}\cdots C_{\alpha_k})\nn\\
    &=\frac{1}{N}\sum_{\alpha_1}\cdots\sum_{\alpha_k}\langle W(C_{\alpha_1}\cdots C_{\alpha_k})\rangle,
\end{align}
where we define $\langle O\rangle$ to be the expectation value of operator $O$ in topological 2D YM:
\begin{align}
    \langle O\rangle\equiv\frac{1}{Z_\YM(0)}\int\c{D}A\,\delta[F]O.
\end{align}

\bigskip

\setcounter{equation}{0}
\renewcommand{\theequation}{B\arabic{equation}}

\sectitle{Appendix B: Peter-Weyl theorem}For a fictitious quantum particle living on the group manifold $U(N)$, the Peter-Weyl theorem states that dual to the ``position'' basis \{$\ket{U}$\}, $U\in U(N)$, there exists a ``momentum'' basis $\{|\c{D}_{ij}^{(R)}\rangle\}$ that is also complete and orthonormal, where $d_R$ is the dimension of irrep $R$ and $1\leq i,j\leq d_R$. The wave functions of $|\c{D}_{ij}^{(R)}\rangle$ in the position basis are the $U(N)$ representation matrices:
\begin{align}
    \c{D}_{ij}^{(R)}(U)=\langle U|\c{D}_{ij}^{(R)}\rangle.
\end{align}
From the completeness relation of the irrep basis,
\begin{align}
\sum_Rd_R\sum_{ij}|\c{D}^{(R)}_{ij}\rangle\langle\c{D}^{(R)}_{ij}|=\hat{\mathds{1}},
\end{align}
formula (\ref{deltaUN}) follows as:
\begin{align}
    \delta(U,V)&=\langle U|V\rangle\nn\\
    &=\sum_R d_R\sum_{ij}\langle U|\c{D}^{(R)}_{ij}\rangle\langle\c{D}^{(R)}_{ij}|V\rangle\nn\\
    &=\sum_R d_R\sum_{ij}\c{D}_{ij}^{(R)}(U)\c{D}_{ji}^{(R)}(V^{-1})\nn\\
    &=\sum_R d_R\chi_R(UV^{-1}),
\end{align}
using the unitarity property $\c{D}^{(R)}(V)^\dag=\c{D}^{(R)}(V^{-1})$. Here, $\chi_R(U)\equiv\tr\c{D}^{(R)}(U)$ is the character of irrep $R$.

\bigskip

\setcounter{equation}{0}
\renewcommand{\theequation}{C\arabic{equation}}

\sectitle{Appendix C: Wilson loops}To evaluate the expectation value $\langle W(C)\rangle$ in the Migdal formulation, we cut the Riemann surface $\Sigma$ along $C$ and introduce link variables $U_\ell$ along this curve. The insertion $W(C)$ is then simply a character $\chi_F\bigl(\prod_{\ell\in C}U_\ell\bigr)$ in the fundamental ($F$) representation:
\begin{align}\label{WC}
    \langle W(C)\rangle=\frac{1}{Z_\YM(0)}\prod_\ell\int dU_\ell\,\chi_F\bigl(\textstyle\prod_{\ell\in C}U_\ell\bigr)\displaystyle\prod_P W_P.
\end{align}
Consider a homologically nontrivial Wilson loop along $C\cong\gamma=\gamma_{\alpha_1}\cdots\gamma_{\alpha_k}\notin[\Gamma,\Gamma]$. Since $\gamma$ is expressed as a word in the generators $\gamma_j$ of $\Gamma$, which correspond to link variables $U_j$ on the boundary segments of the octagon [Fig.~\ref{fig:bolza}(c)], no new plaquette is generated by the Wilson loop insertion and thus Eq.~(\ref{WC}) becomes
\begin{align}\label{homnontrivial1}
    \langle W(C)\rangle&=\frac{1}{Z_\YM(0)}\prod_j\int dU_j\sum_R d_R\chi_F(U_{\alpha_1}\cdots U_{\alpha_k})\nn\\
    &\hspace{15mm}\times\chi_R\bigl(\c{R}(\{U_j\})\bigr).
\end{align}
Using Fubini's theorem~\cite{BtD}, we can write each Haar integral over $U(N)$ as a convolution
\begin{align}
    \int_{U(N)}dU\,f(U)=\int_{PU(N)}dV\int_{U(1)}dW\,f(VW),
\end{align}
where $PU(N)\cong SU(N)/\mathbb{Z}_N$ is the projective unitary group, i.e., the quotient of $U(N)$ by its center $Z(U(N))\cong U(1)=\{e^{i\theta}:\theta\in[0,2\pi)\}$. The generator of this $U(1)$ normal subgroup is the $N\times N$ identity matrix. In the relator $\c{R}(\{U_j\})$, each $U_j$ appears together with its inverse $U_j^{-1}$, thus $\c{R}(\{V_jW_j\})=\c{R}(\{V_j\})$ for $W_j\in U(1)$, while
\begin{align}
    \chi_F(V_{\alpha_1}W_{\alpha_1}\cdots V_{\alpha_k}W_{\alpha_k})=e^{i\sum_jm_j\theta_j}\chi_F(V_{\alpha_1}\cdots V_{\alpha_k}),
\end{align}
where $m_j\in\mathbb{Z}$ is the number of times the generator $\gamma_j$ appears in the word $\gamma_{\alpha_1}\cdots\gamma_{\alpha_k}$ (with $\gamma_j^{-1}$ counted negatively). We thus obtain
\begin{align}\label{HomNontrivialVanish}
    \langle W(C)\rangle&=\frac{1}{Z_\YM(0)}\sum_R d_R\prod_j\int_{PU(N)}dV_j\,
    \chi_F(V_{\alpha_1}\cdots V_{\alpha_k})\nn\\
    &\hspace{15mm}\times\chi_R\bigl(\c{R}(\{V_j\})\bigr)
    \prod_j\int_{U(1)}\frac{d\theta_j}{2\pi}e^{i\sum_jm_j\theta_j}\nn\\
    &\propto\prod_j\delta_{m_j,0}.
\end{align}
Since $\gamma\notin[\Gamma,\Gamma]$ by assumption, there is a mismatch between the number of generators and their inverses appearing in the word $\gamma_{\alpha_1}\cdots\gamma_{\alpha_k}$. Thus at least one of the $m_j$ is nonzero, and $\langle W(C)\rangle$ vanishes.

Finally, homologically trivial noncontractible loops are those for which $\gamma\in[\Gamma,\Gamma]$ but $\gamma\neq 1$. In Ref.~\cite{magee2021}, Magee considers the function $\tr_\gamma:\c{M}_{g,N}\rightarrow\mathbb{C}$ defined by
\begin{align}
    \tr_\gamma(\phi)\equiv\tr\phi(\gamma),
\end{align}
and its expectation value
\begin{align}\label{magee1}
    \mathbb{E}_{g,N}[\tr_\gamma]\equiv\frac{\int_{\c{M}_{g,N}}\tr_\gamma\,d\mathrm{Vol}_{\c{M}_{g,N}}}
    {\int_{\c{M}_{g,N}}\,d\mathrm{Vol}_{\c{M}_{g,N}}}.
\end{align}
Here, $\c{M}_{g,N}\equiv\hom(\Gamma_g,SU(N))_\mathrm{irr}/PSU(N)$ is the moduli space of (irreducible) flat $SU(N)$ connections on a Riemann surface $\Sigma$ of genus $g$ with fundamental group $\Gamma_g$, up to gauge equivalence; $\tr_\gamma$ is the trace in the fundamental representation of the $SU(N)$ holonomy around loop $\gamma\in\Gamma_g$; and $d\mathrm{Vol}_{\c{M}_{g,N}}$ is the Atiyah-Bott-Goldman measure on $\c{M}_{g,N}$. Thus, Eq.~(\ref{magee1}) is nothing but the expectation value of an $SU(N)$ Wilson loop around loop $C\cong\gamma$ in 2D topological YM theory; note that any prefactor ambiguities in the definition of the measure cancel out in the expectation value. For loops $\gamma=\gamma_{\alpha_1}\cdots\gamma_{\alpha_k}$ such that $\gamma\in[\Gamma,\Gamma]$, $U(N)$ can be replaced by $SU(N)$ without changing the expectation value, because in both the holonomy $U_{\alpha_1}\cdots U_{\alpha_k}$ and the relator $\c{R}(\{U_j\})$ in the general expression (\ref{homnontrivial1}), each generator $U_j$ appears together with its inverse, and the $U(1)$ part of those $U_j$'s cancels.

Ref.~\cite{magee2021} investigates the behavior of Eq.~(\ref{magee1}) in the large-$N$ limit. Applied to the present context, Magee proves using Weingarten calculus~\cite{SM} that
\begin{align}
\langle W(C)\rangle=\mathbb{E}_{g,N}[\tr_\gamma]\sim\c{O}(1)\text{ as }N\rightarrow\infty,\text{ for }\gamma\neq 1.
\end{align}
Having already shown in Eq.~(\ref{HomNontrivialVanish}) that homologically nontrivial $U(N)$ Wilson loops vanish identically for all $N$, this implies that, in the large-$N$ limit, homologically trivial noncontractible loops $\gamma\in[\Gamma,\Gamma]$, $\gamma\neq 1$ give a contribution to the DOS moments (\ref{moment-largeN}) that is smaller by a factor of $1/N$ than those of contractible loops (\ref{moment-infinity}). Accordingly, in the limit $N\rightarrow\infty$, only contractible loops contribute, which leads to our final result.

\bibliography{main}

\begin{thebibliography}{91}
\bibitem [{rot()}]{rotor}%
  \BibitemOpen
  \href@noop {} {}\bibinfo {note} {As a more familiar analogy, the
  eigenfunctions of a quantum rigid-body rotor---a quantum particle ``living''
  on the group $SO(3)$---are the Wigner $D$-matrices $D_{mm'}^{(J)}(R)^*$, with
  $m,m'=-J,\ldots,J$~\cite{LL-QM}.}\BibitemShut {Stop}%
\end{thebibliography}
\begin{thebibliography}{30}
\bibitem [{tra()}]{transpose}%
  \BibitemOpen
  \href@noop {} {}\bibinfo {note} {Note that it is the transpose
  $D^{(K)}(\gamma_\alpha)^T$ which appears in the matrix element
  (\ref{Hblock-diagonal}), but this does not affect spectral properties such as
  $\rho_N(E)$.}\BibitemShut {Stop}%
\bibitem [{irr()}]{irreps}%
  \BibitemOpen
  \href@noop {} {}\bibinfo {note} {Although both play an important role in this
  work, the discrete set $\{R\}$ of irreps of the compact Lie group $U(N)$,
  with irrep-dependent dimensions $d_R$, is not to be confused with the
  continuous set $\{K\}\equiv\BZ$ of $N$-dimensional irreps of the infinite,
  discrete Fuchsian group $\Gamma$. Accordingly, the matrices $\c{D}^{(R)}(U)$
  for the former and $D^{(K)}(\gamma)$ for the latter are different
  objects.}\BibitemShut {Stop}%
\end{thebibliography}

\end{document}


\newcommand{\bsigma}{\boldsymbol{\sigma}}
\newcommand{\re}{\mathop{\mathrm{Re}}}
\newcommand{\im}{\mathop{\mathrm{Im}}}
\renewcommand{\b}[1]{{\boldsymbol{#1}}}
\newcommand{\diag}{\mathrm{diag}}
\newcommand{\sign}{\mathrm{sign}}
\newcommand{\sgn}{\mathop{\mathrm{sgn}}}
\renewcommand{\c}[1]{\mathcal{#1}}
\newcommand{\vol}{\mathop{\mathrm{vol}}}
\newcommand{\BZ}{{\mathrm{BZ}_N}}
\newcommand{\YM}{{\textrm{YM}}}

\newcommand{\mb}{\bm}
\newcommand{\ua}{\uparrow}
\newcommand{\da}{\downarrow}
\newcommand{\ra}{\rightarrow}
\newcommand{\la}{\leftarrow}
\newcommand{\mc}{\mathcal}
\newcommand{\bs}{\boldsymbol}
\newcommand{\lra}{\leftrightarrow}
\newcommand{\nn}{\nonumber}
\newcommand{\half}{{\textstyle{\frac{1}{2}}}}
\newcommand{\mf}{\mathfrak}
\newcommand{\MF}{\text{MF}}
\newcommand{\IR}{\text{IR}}
\newcommand{\UV}{\text{UV}}
\newcommand{\so}{\mathfrak{so}}
\renewcommand{\geq}{\geqslant}
\renewcommand{\leq}{\leqslant}
\renewcommand{\hom}{{\mathrm{Hom}}}
\renewcommand{\d}{{\mathrm{d}}}

\title{Supplemental Material for:\texorpdfstring{\\}{ }Hyperbolic lattices and two-dimensional Yang-Mills theory}

\author{G. Shankar}
\email{sankaran@ualberta.ca}
\affiliation{Department of Physics, University of Alberta, Edmonton, Alberta T6G 2E1, Canada}

\author{Joseph Maciejko\,\orcidlink{0000-0002-6946-1492}}
\email{maciejko@ualberta.ca}
\affiliation{Department of Physics, University of Alberta, Edmonton, Alberta T6G 2E1, Canada}
\affiliation{Theoretical Physics Institute \& Quantum Horizons Alberta, University of Alberta, Edmonton, Alberta T6G 2E1, Canada}

\date{\today}

\maketitle

\tableofcontents

\section{Hyperbolic band theory}

In this section, we review and streamline the formulation of hyperbolic band theory~\cite{maciejko2021,maciejko2022}, introducing the notions of non-Abelian Bloch states and non-Abelian Bloch Hamiltonians for a generic $\{p,q\}$ lattice, that generalize the discussion of the $\{8,8\}$ lattice in the main text.

A hyperbolic $\{p,q\}$ lattice can be represented as a collection of Bravais unit cells together with a finite sublattice basis of $n$ sites per cell~\cite{boettcher2022}. The fact that this is true for any hyperbolic $\{p,q\}$ lattice follows from the (now proven) Fenchel's conjecture, according to which the space group $\Delta$ of the $\{p,q\}$ lattice admits a (normal) translation subgroup $\Gamma$ of finite index~\cite{chen2023}. The infinite Fuchsian group $\Gamma$ is isomorphic to the fundamental group $\pi_1(\Sigma)$ of a smooth, compact Riemann surface $\Sigma$ of genus $g\geq 2$ that results from identification of the sides of the Bravais unit cell under the action of $\Gamma$. The finite index $|\Delta\colon\Gamma|$ of $\Gamma$ in $\Delta$ is equal to the order of the point group $G=\Delta/\Gamma$. A given vertex $x_0$ of the $\{p,q\}$ lattice is invariant under its site-symmetry group or stabilizer subgroup $G_{x_0}\subset G$, and by the orbit-stabilizer theorem, the set of all vertices contained in the Bravais unit cell (i.e., the orbit $G(x_0)$ of $x_0$) is in one-to-one correspondence with the elements of the coset space $G/G_{x_0}$~\cite{bzdusek2022,lenggenhager2023}. Since $G$ is finite, $G_{x_0}$ is also finite, and by Lagrange's theorem, $n\equiv|G/G_{x_0}|$ is finite.

\subsection{Real space: the Fuchsian group basis}

A complete, orthonormal basis of Hilbert space for a single quantum particle hopping on a $\{p,q\}$ lattice is the position basis $\{\ket{\gamma}\otimes\ket{\sigma}\}$, where $\gamma\in\Gamma$ labels Fuchsian group elements and $\sigma$ collectively labels a finite set of intracell degrees of freedom like sublattice, orbital, and/or spin degrees of freedom. We are interested in the spectrum of $\Gamma$-invariant finite-range hopping Hamiltonians, of the form
\begin{align}\label{H}
    \hat{H}=\sum_{\gamma\in\Gamma}\sum_\alpha\ket{\gamma\gamma_\alpha}\bra{\gamma}\otimes T(\gamma_\alpha),
\end{align}
where $\alpha$ indexes a finite set $\{\gamma_\alpha\}$ of $\Gamma$-translations in which each element appears together with its inverse, and Hermiticity of $\hat{H}$ requires that the finite-dimensional hopping matrix $T(\gamma_\alpha)$ obey $T(\gamma_\alpha^{-1})=T^\dag(\gamma_\alpha)$. The set $\{\gamma_\alpha\}$ can always be expressed in terms of the identity and the generators $\gamma_j$, $j=1,\ldots,2g$ of $\Gamma$ and their inverses, which obey a single relation $R(\{\gamma_j\})=1$, in which a product of $4g$ generators involving each $\gamma_j$ and $\gamma_j^{-1}$ once is set to the identity $1\in\Gamma$. For the nearest-neighbor model on the $\{8,8\}$ lattice considered in the main text, $\{\gamma_\alpha\}=\{\gamma_1,\gamma_2,\gamma_3,\gamma_4\}$ and the relator $R$ is $R(\{\gamma_j\})=\gamma_1\gamma_2^{-1}\gamma_3\gamma_4^{-1}\gamma_1^{-1}\gamma_2\gamma_3^{-1}\gamma_4$ as mentioned in the main text. The hopping matrices $T(\gamma_\alpha)$ are trivial in this example due to the lack of sublattice/orbital/spin degrees of freedom.

\begin{figure}[t!]
\centering
    \includegraphics[width=0.5\columnwidth]{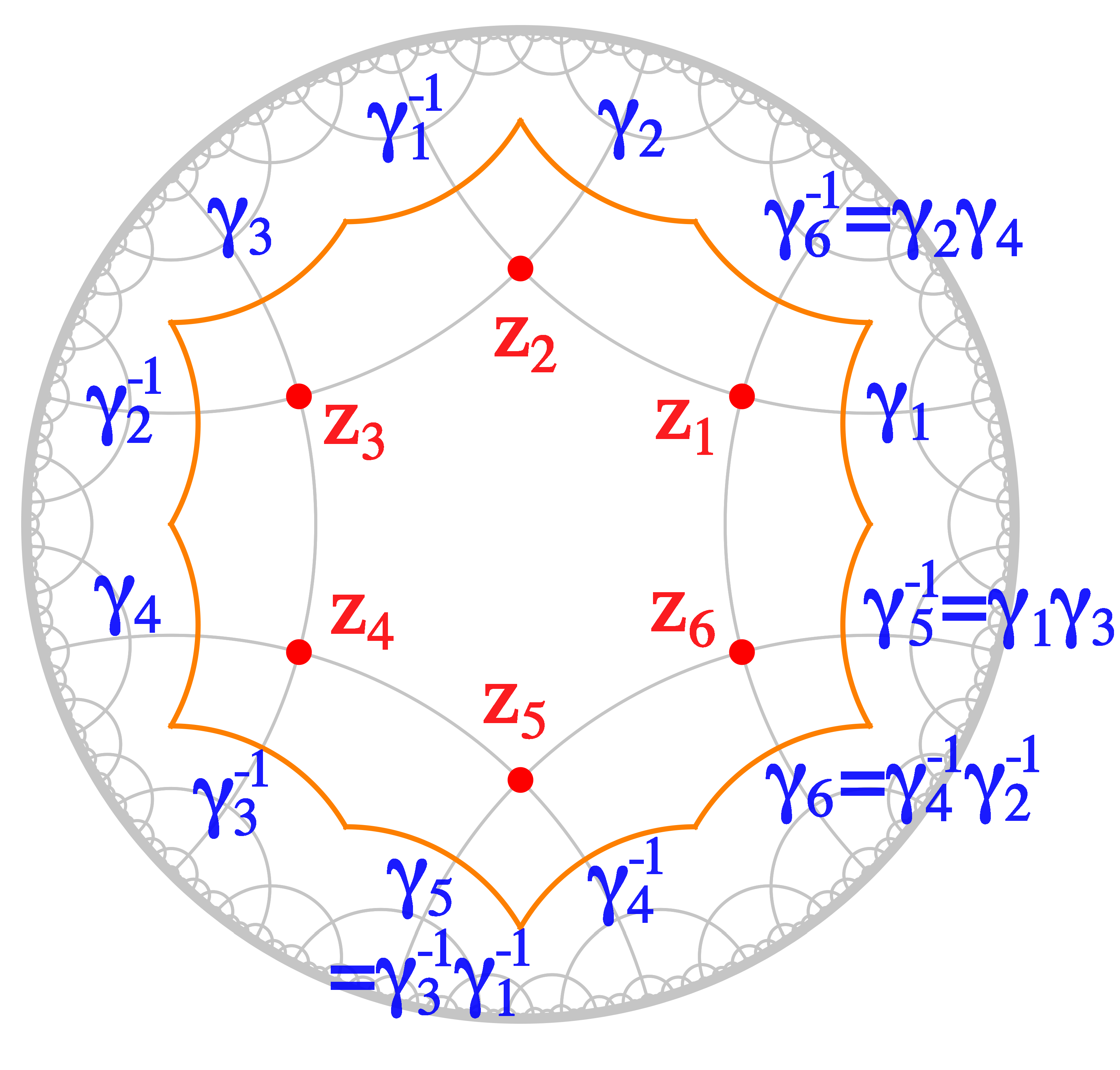}
    \caption[]{The $\{6,4\}$ lattice (light grey) can be described as a hyperbolic Bravais lattice with a 12-sided unit cell (orange) which contains 6 sites (red) and compactifies onto a Riemann surface of genus $g=2$.}
\label{fig:64}
\end{figure}

A slightly more complicated example is the $\{6,4\}$ lattice (Fig.~\ref{fig:64}), which can be decomposed into a Bravais lattice decorated with 6 sites per dodecagonal unit cell~\cite{chen2023}. Such a lattice has six translation generators $\{\gamma_1,\ldots,\gamma_6\}$ obeying three relational constraints: $\gamma_1\gamma_3\gamma_5\!=\!1$, $\gamma_2\gamma_4\gamma_6\!=\!1$, and $\gamma_1\gamma_2\gamma_3\gamma_4\gamma_5\gamma_6\!=\!1$. Using the first two relations, one can eliminate $\gamma_5\!=\!\gamma_3^{-1}\gamma_1^{-1}$ and $\gamma_6\!=\!\gamma_4^{-1}\gamma_2^{-1}$ in favor of a reduced set of four generators $\gamma_1,\gamma_2,\gamma_3,\gamma_4$ obeying a single relation: $R(\{\gamma_j\})=\gamma_1\gamma_2\gamma_3\gamma_4\gamma_3^{-1}\gamma_1^{-1}\gamma_4^{-1}\gamma_2^{-1}\!=\!1$. For a nearest-neighbor hopping model on such a lattice, the summation over $\alpha$ in Eq.~\eqref{H} is one over $\gamma_\alpha\!\in\!\{1,\gamma_1,\ldots,\gamma_6,\gamma_1^{-1},\ldots, \gamma_6^{-1}\}$, where once again $\gamma_5,\gamma_6$ and their inverses can be replaced by their equivalent expressions in terms of $\gamma_1,\dots,\gamma_4$ and their inverses. The $6\!\times\!6$ hopping matrices $T(1)$ and $T(\gamma_j)$ describe intra-cell and inter-cell hopping respectively. Indexing the sites within a unit cell as ${z_1,z_2,...,z_6}$ and $z_7\!\equiv\!z_1$, the hopping matrices can be explicitly written as 
\begin{align}
T(1) &= \sum_{i=1}^{6} \left(\dyad{z_i}{z_{i+1}}+\dyad{z_{i+1}}{z_{i}}\right),\nonumber \\
T(\gamma_i) &= \dyad{z_i}{z_{i+1}},\quad T(\gamma_i^{-1}) = T(\gamma_i)^\dagger, \quad i\!\in\!\{1,2,3,4,6\}.
\end{align}
Hopping terms beyond nearest neighbor, for example second-neighbor hopping terms as in the Haldane model on the $\{6,4\}$ lattice (see Fig.~1 in Ref.~\cite{chen2023}), can be straightforwardly included by enlarging the set $\{\gamma_\alpha\}$ to include second-neighbor translations and defining the corresponding $6\times 6$ hopping matrices.

More generally, there always exists a canonical choice of generators $\{\gamma_1,\ldots,\gamma_g;\gamma_{g+1},\ldots,\gamma_{2g}\}=\{a_1,\ldots,a_g;b_1,\ldots,b_g\}$ such that the relation among them can be expressed as
\begin{align}
    R(\{a_j,b_j\})=[a_1,b_1][a_2,b_2]\cdots[a_g,b_g]=1,
\end{align}
where $[a,b]\equiv aba^{-1}b^{-1}$ is the commutator of $a$ and $b$~\cite{FarkasKra}.

As a consequence of translation invariance, the Hamiltonian (\ref{H}) commutes with the unitary translation operators
\begin{align}
    \hat{L}_\gamma=\sum_{\gamma'\in\Gamma}\ket{\gamma\gamma'}\bra{\gamma'},\hspace{5mm}\gamma\in\Gamma,
\end{align}
which obey $\hat{L}^\dag_{\gamma}=\hat{L}_{\gamma^{-1}}$, and act on position eigenkets as $\hat{L}_\gamma\ket{\gamma_0}=\ket{\gamma\gamma_0}$, i.e., according to the (left-)regular representation~\cite{FultonHarris}. Note that we omit identity operators in sublattice/orbital/spin space when it is clear they are implicitly tensored. By translation symmetry, we thus expect eigenstates of $\hat{H}$ to transform under $\hat{L}_\gamma$ according to unitary irreps $K$ of $\Gamma$.

\subsection{Reciprocal space: the non-Abelian Bloch Hamiltonian}

For each $N\geq 1$, there exists a moduli space of $N$-dimensional irreps $K\in\BZ$ with unitary representation matrices $D_{\nu\lambda}^{(K)}(\gamma)=D_{\lambda\nu}^{(K)}(\gamma^{-1})^*$. For fixed $N$ and $K\in\BZ$, we use these to construct the $N^2$ basis states
\begin{align}\label{basis}
    \ket{K,\lambda,\nu}\equiv\sum_{\gamma\in\Gamma}D_{\nu\lambda}^{(K)}(\gamma)^*\ket{\gamma},\hspace{5mm}\lambda,\nu=1,\ldots,N,
\end{align}
which transform under $\hat{L}_\gamma$ as
\begin{align}
    \hat{L}_{\gamma'}\ket{K,\lambda,\nu}&=\sum_{\gamma\in\Gamma}D_{\nu\lambda}^{(K)}(\gamma)^*\ket{\gamma'\gamma}\nn\\
    &=\sum_{\gamma\in\Gamma}D_{\nu\lambda}^{(K)}(\gamma^{\prime -1}\gamma)^*\ket{\gamma}\nn\\
    &=\sum_{\gamma\in\Gamma}\sum_\mu D_{\nu\mu}^{(K)}(\gamma^{\prime -1})^*D_{\mu\lambda}^{(K)}(\gamma)^*\ket{\gamma}\nn\\
    &=\sum_\mu\ket{K,\lambda,\mu}D_{\mu\nu}^{(K)}(\gamma'),
\end{align}
i.e., according to the non-Abelian Bloch theorem~\cite{maciejko2022}. For fixed $\lambda$, the $N$ states $\{\ket{K,\lambda,\nu},\,\nu=1,\ldots,N\}$ form a degenerate multiplet that transforms into itself under translations. As discussed in Appendix A of the main text, by approximating $\Gamma$ as an arbitrarily large but finite group of order $|\Gamma|\gg 1$, e.g., as a quotient $\Gamma/\Gamma'$ by a normal subgroup $\Gamma'$ of large index~\cite{maciejko2022, lux2022,lenggenhager2023}, one can argue that the basis states (\ref{basis}) must obey a Schur-type orthogonality relation:
\begin{align}
    \braket{K,\lambda,\nu}{K',\lambda',\nu'}=\delta(K-K')\delta_{\lambda\lambda'}\delta_{\nu\nu'}.
\end{align}

Finally, we show that in the subspace spanned by $\{\ket{K,\lambda,\nu,\sigma}=\ket{K,\lambda,\nu}\otimes\ket{\sigma}\}$, the Hamitonian (\ref{H}) is block-diagonal:
\begin{align}\label{Hblock-diagonal}
    \langle K,\lambda,\nu,\sigma|\hat{H}|K',\lambda',\nu',\sigma'\rangle&=\sum_{\gamma\in\Gamma}\sum_{\alpha}\langle K,\lambda,\nu|\gamma\gamma_\alpha\rangle\langle\gamma|K',\lambda',\nu'\rangle T_{\sigma\sigma'}(\gamma_\alpha)\nn\\
&=\sum_{\gamma\in\Gamma}\sum_{\alpha} D_{\nu\lambda}^{(K)}(\gamma\gamma_\alpha)D_{\nu'\lambda'}^{(K')}(\gamma)^*T_{\sigma\sigma'}(\gamma_\alpha)\nn\\
&=\sum_{\gamma\in\Gamma}\sum_{\alpha}\sum_\mu D_{\nu\mu}^{(K)}(\gamma)D_{\mu\lambda}^{(K)}(\gamma_\alpha)D_{\nu'\lambda'}^{(K')}(\gamma)^*T_{\sigma\sigma'}(\gamma_\alpha)\nn\\
&=\sum_{\alpha}\sum_\mu D_{\mu\lambda}^{(K)}(\gamma_\alpha)T_{\sigma\sigma'}(\gamma_\alpha)\braket{K,\mu,\nu}{K',\lambda',\nu'}\nn\\
&=\delta(K-K')\delta_{\nu\nu'}\sum_{\alpha} D_{\lambda'\lambda}^{(K)}(\gamma_\alpha)T_{\sigma\sigma'}(\gamma_\alpha)\nn\\
&=\delta(K-K')\delta_{\nu\nu'}\c{H}_{\lambda'\lambda}^{\sigma\sigma'}(K),
\end{align}
where
\begin{align}
    \c{H}(K)=\sum_\alpha D^{(K)}(\gamma_\alpha)\otimes T(\gamma_\alpha),
\end{align}
is the $U(N)$ Bloch Hamiltonian. In the main text, we discuss the simplest situation where there is a single degree of freedom per Bravais unit cell, thus $T(\gamma_\alpha)=1$ and $\c{H}(K)$ is a $N\times N$ matrix. In general, if there are $d$ intracell degrees of freedom, the Bloch Hamiltonian is a $Nd\times Nd$ matrix.

\subsection{Moments of the $U(N)$ density of states}

In the main text, we define the $U(N)$ density of states as the density of states of the $U(N)$ Bloch Hamiltonian for fixed $N$~\cite{transpose}:
\begin{align}
    \rho_N(E)\equiv\frac{1}{N}\int_\BZ\frac{dK}{\vol(\BZ)}\tr\delta(E-\c{H}(K)).
\end{align}
We compute the corresponding $k$th moment as
\begin{align}\label{UNmoment}
    \rho_N^{(k)}&=\int dE\,\rho_N(E)E^k\nn\\
    &=\frac{1}{N}\int_\BZ\frac{dK}{\vol(\BZ)}\int dE\,E^k\tr\delta(E-\c{H}(K))\nn\\
    &=\frac{1}{N}\int_\BZ\frac{dK}{\vol(\BZ)}\tr\c{H}(K)^k\nn\\
    &=\frac{1}{N}\int_\BZ\frac{dK}{\vol(\BZ)}\sum_{\alpha_1}\cdots\sum_{\alpha_k}\tr D^{(K)}(\gamma_{\alpha_1})\cdots D^{(K)}(\gamma_{\alpha_k})\cdot\tr T(\gamma_{\alpha_1})\cdots T(\gamma_{\alpha_k})\nn\\
    &=\frac{1}{N}\int_\BZ\frac{dK}{\vol(\BZ)}\sum_{\alpha_1}\cdots\sum_{\alpha_k}\tr D^{(K)}(\gamma_{\alpha_1}\cdots\gamma_{\alpha_k})\cdot\tr T(\gamma_{\alpha_1})\cdots T(\gamma_{\alpha_k}).
\end{align}
In the last line, we have used the representation property of $D^{(K)}$, i.e., $D^{(K)}(\gamma)D^{(K)}(\gamma')=D^{(K)}(\gamma\gamma')$. The various terms in the sum (\ref{UNmoment}) factorize into a piece containing information about the geometry of the Bravais lattice ($\tr D^{(K)}$) and a piece describing the structure of hopping in sublattice/orbital/spin space ($\tr T\cdots T$). In the main text, we consider the case $T(\gamma_\alpha)=1$ and thus $\tr T(\gamma_{\alpha_1})\cdots T(\gamma_{\alpha_k})=1$.

\section{2D Yang-Mills theory}

In this section, we summarize aspects of 2D Yang-Mills theory~\cite{migdal1975,witten1991,cordes1995} that are relevant in the context of hyperbolic lattices, and provide further background and context to understand and appreciate our results. In Sec.~\ref{sec:migdal} and \ref{sec:subdivision}, we show that Migdal's lattice gauge theory is equivalent to continuum 2D Yang-Mills theory. In Sec.~\ref{sec:WL}, we give further details regarding the calculation of Wilson-loop expectation values.

\subsection{Migdal's lattice gauge theory}\label{sec:migdal}

We begin with the Hamiltonian of continuum Yang-Mills theory in 1+1 dimensions, i.e., pure 2D quantum chromodynamics (QCD$_2$) without matter:
\begin{align}
\hat{H}_\YM=\frac{e^2}{2}\int dx\,\hat{E}^A_x\hat{E}^A_x,
\end{align}
where the sum over the repeated indices $A$ labeling generators of the gauge group $U(N)$ is understood, and $\hat{E}^A_x(x)$ are continuum electric field operators. The Euclidean partition function is
\begin{align}
Z_\text{YM}=\tr e^{-\beta\hat{H}_\YM},
\end{align}
where the trace is only over gauge-invariant states. We first place this theory on a discrete lattice~\cite{Smit}. For simplicity, we consider a rectangular $L\times L_\tau$ spacetime lattice, with lattice constant $a$ in the spatial direction $x$ and $\epsilon$ in the imaginary time direction $\tau$. We thus have a 1D lattice Hamiltonian
\begin{align}
\hat{H}_\text{lat}=\frac{e^2a}{2}\sum_{x=1}^L\hat{E}_{x,x+1}^A\hat{E}_{x,x+1}^A,
\end{align}
where the lattice electric fields on links $(x,x+1)$ and $(x',x'+1)$ satisfy the commutation relations
\begin{align}
[\hat{E}_{x,x+1}^A,\hat{E}_{x',x'+1}^B]=if^{ABC}\hat{E}_{x,x+1}^C\delta_{xx'},
\end{align}
with $f^{ABC}$ the structure constants of $U(N)$. Cutting the partition function into $L_\tau\rightarrow\infty$ Suzuki-Trotter time slices, we obtain
\begin{align}
Z_\text{YM}=\tr e^{-\epsilon\hat{H}_\text{lat}}e^{-\epsilon\hat{H}_\text{lat}}\cdots e^{-\epsilon\hat{H}_\text{lat}},
\end{align}
with $\epsilon=\beta/L_\tau\rightarrow 0$. We evaluate this trace in the ``position basis''
\begin{align}
\ket{\{U\}}=\bigotimes_{x=1}^L\ket{U_{x,x+1}},
\end{align}
where $U_{x,x+1}$ is an abstract $U(N)$ group element on link $(x,x+1)$. Introducing $L_\tau$ copies of the resolution of the identity in this basis,
\begin{align}
\prod_{x=1}^L\int dU_{x,x+1}(\tau)\ket{U_{x,x+1}(\tau)}\bra{U_{x,x+1}(\tau)}=\hat{\mathds{1}},
\hspace{5mm}\tau=1,\ldots,L_\tau,
\end{align}
where the integral is over the $U(N)$ group manifold with the Haar measure, we have
\begin{align}\label{ZLGT1}
Z_\text{YM}&=\prod_{x=1}^L\prod_{\tau=1}^{L_\tau}\int dU_{x,x+1}(\tau)\bra{U_{x,x+1}(\tau)}e^{-\epsilon\hat{H}_\text{lat}}\ket{U_{x,x+1}(\tau+1)}\nn\\
&=\prod_{x,\tau}\int dU_{x,x+1}(\tau)\bra{U_{x,x+1}(\tau)}\exp\left(-\frac{e^2a\epsilon}{2}\hat{E}_{x,x+1}^A\hat{E}_{x,x+1}^A\right)\ket{U_{x,x+1}(\tau+1)}.
\end{align}
To evaluate the matrix element, we introduce a resolution of the identity in the ``momentum basis'' $\left\{\ket{\c{D}^{(R)}_{ij}}\right\}$, corresponding to irreps $R$ of the gauge group $U(N)$~\cite{irreps}:
\begin{align}
\sum_Rd_R\sum_{ij}\ket{\c{D}^{(R)}_{ij}}\bra{\c{D}^{(R)}_{ij}}=\hat{\mathds{1}},
\end{align}
where $d_R$ is the dimension of irrep $R$ and $1\leq i,j\leq d_R$ (see also Appendix B in the main text). The completeness of the irrep basis for a compact Lie group like $U(N)$ is known as the Peter-Weyl theorem~\cite{BtD}. The representation matrices of $U(N)$ are simply the wave functions of $\ket{\c{D}^{(R)}_{ij}}$ in the position basis:
\begin{align}
\c{D}^{(R)}_{ij}(U)=\braket{U}{\c{D}^{(R)}_{ij}},\hspace{5mm}U\in U(N).
\end{align}
Using the fact that the states $\ket{\c{D}^{(R)}_{ij}}$ are eigenstates of the $U(N)$ quadratic Casimir operator,
\begin{align}
\hat{E}^A\hat{E}^A\ket{\c{D}^{(R)}_{ij}}=c_2(R)\ket{\c{D}^{(R)}_{ij}},
\end{align}
where $c_2(R)$ is a representation-dependent constant, we obtain:
\begin{align}\label{matrixel}
\bra{U}e^{-\alpha\hat{E}^A\hat{E}^A}\ket{U'}&=\sum_R d_R\sum_{ij}\langle U|e^{-\alpha\hat{E}^A\hat{E}^A}|\c{D}^{(R)}_{ij}\rangle\langle\c{D}^{(R)}_{ij}|U'\rangle\nn\\
&=\sum_Rd_Re^{-\alpha c_2(R)}\sum_{ij}\c{D}^{(R)}_{ij}(U)\c{D}^{(R)}_{ij}(U')^*\nn\\
&=\sum_Rd_Re^{-\alpha c_2(R)}\sum_{ij}\c{D}^{(R)}_{ij}(U)\c{D}^{(R)}_{ji}(U'^{-1})\nn\\
&=\sum_Rd_R\chi_R(U'^{-1}U)e^{-\alpha c_2(R)},
\end{align}
where $\chi_R(U)=\tr\c{D}^{(R)}(U)$ is the character of representation $R$. Equation~(\ref{ZLGT1}) thus becomes:
\begin{align}\label{ZYM}
Z_\text{YM}=\prod_{x,\tau}\int dU_{x,x+1}(\tau)\sum_Rd_R\chi_R\left(U_{x,x+1}(\tau+1)^{-1}U_{x,x+1}(\tau)\right)e^{-e^2a\epsilon c_2(R)/2}.
\end{align}
On the other hand, we consider Migdal's lattice gauge theory~\cite{migdal1975}:
\begin{align}\label{Zmigdal}
Z_\text{LGT}=\prod_\ell\int dU_\ell\prod_PW_P,
\end{align}
where $P$ denotes plaquettes and $\ell$ links in 2D spacetime. The plaquette Boltzmann weight is defined as
\begin{align}
W_P=\sum_Rd_R\chi_R\bigl(\textstyle\prod_{\ell\in P}U_\ell\bigr)e^{-e^2A_Pc_2(R)/2},
\end{align}
where $A_P$ is the area of plaquette $P$. We consider rectangular plaquettes of length $a$ in the $x$ direction and $\epsilon$ in the $\tau$ direction. The partition function can then be written as
\begin{align}\label{ZLGT2}
Z_\text{LGT}=\prod_{x,\tau}\prod_{\hat{\mu}}\int dU_{(x,\tau)+\hat{\mu}}\sum_R d_R\chi_R\left(U_{(x,\tau)+\hat{x}}U_{(x+1,\tau)+\hat{\tau}}U_{(x+1,\tau+1)-\hat{x}}U_{(x,\tau+1)-\hat{\tau}}\right)e^{-e^2a\epsilon c_2(R)/2},
\end{align}
where $\hat{\mu}=\hat{x},\hat{\tau}$ denote the two unit vectors on the spacetime lattice. We now use the gauge invariance of the partition function to work in the temporal gauge, where $U$ can be set to identity on all temporal links: $U_{(x,\tau)+\hat{\tau}}=1$, for all $x,\tau$. Using $U_{ji}=U_{ij}^{-1}$, and denoting $U_{(x,\tau)+\hat{x}}\equiv U_{x,x+1}(\tau)$, Eq.~(\ref{ZLGT2}) then becomes:
\begin{align}
Z_\text{LGT}=\prod_{x,\tau}\int dU_{x,x+1}(\tau)\sum_R d_R\chi_R\left(U_{x,x+1}(\tau)U_{x,x+1}(\tau+1)^{-1}\right)e^{-e^2a\epsilon c_2(R)/2}=Z_\text{YM},
\end{align}
upon comparing with Eq.~(\ref{ZYM}). Thus Migdal's lattice gauge theory (\ref{Zmigdal}) reduces to latticized 2D Yang-Mills theory. In the topological limit $e^2=0$, the plaquette Boltzmann weight is
\begin{align}\label{WPtop}
    W_P=\sum_Rd_R\chi_R\bigl(\textstyle\prod_{\ell\in P}U_\ell\bigr).
\end{align}

\begin{figure}[t!]
\centering
    \includegraphics[width=0.75\columnwidth]{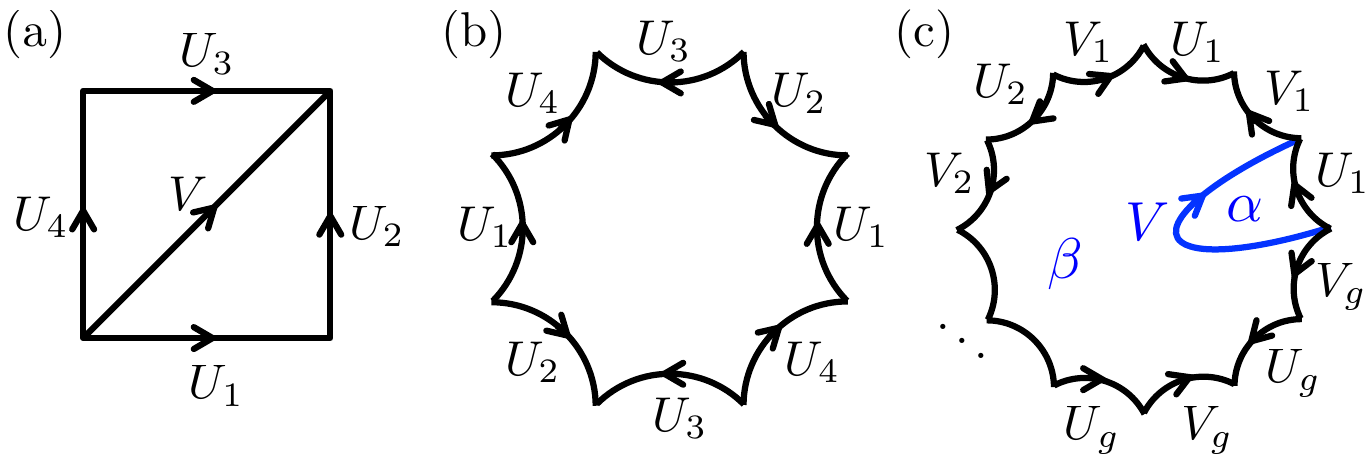}
    \caption[]{Migdal's lattice gauge theory. (a) The partition function remains invariant under changes of discretization, e.g., upon adding to the square a diagonal link. Subdivision invariance allows one to simplify calculations in 2D Yang-Mills theory by using the coarsest possible discretization of a genus-$g$ surface $\Sigma$, i.e., a polygon with $4g$ sides pairwise identified and $2g$ link variables. (b) Discretization of the genus-2 Bolza surface~\cite{maciejko2021} relevant for the $\{8,8\}$ lattice considered in the main text; (c) discretization (in black) of a generic genus-$g$ surface. Blue: additional curve on $\Sigma$ used in the example Wilson loop computation in Sec.~\ref{sec:WL}.}
\label{fig:migdal}
\end{figure}

\subsection{Subdivision invariance}\label{sec:subdivision}

In 2D, Migdal's lattice gauge theory exhibits the property of subdivision invariance: the partition function (\ref{Zmigdal}) is independent of the choice of discretization of the spacetime surface~\cite{migdal1975,witten1991}. To see this, consider the partition function on a single square $P$ of area $A$ [Fig.~\ref{fig:migdal}(a)]. We first discretize $P$ with four links on its perimeter: the partition function is then
\begin{align}\label{Zsquare}
    Z_\text{LGT}=\int dU_1dU_2dU_3dU_4\,W_P,
\end{align}
where the Boltzmann weight is
\begin{align}
    W_P=\sum_R d_R\chi_R(U_1U_2U_3^{-1}U_4^{-1})e^{-e^2Ac_2(R)/2},
\end{align}
recalling the fact that when traversing a link in reverse order, one must take the inverse of the link group variable~\cite{Smit}. Now consider refining the discretization of $P$ by introducing an additional diagonal link with corresponding link variable $V$ [Fig.~\ref{fig:migdal}(a)]. $P$ now contains two triangular plaquettes $P'$ and $P''$, such that the partition function is
\begin{align}\label{ZLGTexample}
    \tilde{Z}_\text{LGT}=\int dU_1dU_2dU_3dU_4dV\,W_{P'}W_{P''},
\end{align}
with the Boltzmann weights
\begin{align}
    W_{P'}&=\sum_{R'}d_{R'}\chi_{R'}(U_1U_2V^{-1})e^{-e^2A'c_2(R')/2},\\
    W_{P''}&=\sum_{R''}d_{R''}\chi_{R''}(VU_3^{-1}U_4^{-1})e^{-e^2A''c_2(R'')/2},
\end{align}
where $A'$ and $A''$ are the areas of $P'$ and $P''$. Performing the $V$ integral in Eq.~(\ref{ZLGTexample}), we have
\begin{align}\label{subdiv}
    \int dV\,W_{P'}W_{P''}&=\sum_{R'R''}d_{R'}d_{R''}e^{-e^2[A'c_2(R')+A''c_2(R'')]/2}
    \int dV\,\chi_{R'}(U_1U_2V^{-1})\chi_{R''}(VU_3^{-1}U_4^{-1})\nn\\
    &=\sum_R d_R e^{-e^2(A'+A'')c_2(R)/2}\chi_R(U_1U_2U_3^{-1}U_4^{-1})\nn\\
    &=W_P,
\end{align}
where we have used $A'+A''=A$ as well as the following identity that follows from the orthonormality of $U(N)$ characters~\cite{BtD}:
\begin{align}\label{ortho1}
    \int dV\,\chi_{R'}(AV^{-1})\chi_{R''}(VB)=\delta_{R'R''}\frac{1}{d_R'}\chi_{R'}(AB).
\end{align}
From Eq.~(\ref{subdiv}), we conclude that the partition function (\ref{ZLGTexample}) with finer discretization is indeed equal to the partition function (\ref{Zsquare}) with coarser discretization. This has two important consequences. On the one hand, one can refine the discretization {\it ad infinitum} without changing the partition function. In the limit of infinite discretization, we recover the continuum limit: thus the lattice partition function (\ref{Zmigdal}) is {\it exactly} equal to the continuum partition function $Z_\YM(e^2)$, with suitably chosen regularization. On the other hand, one can also maximally coarsen the discretization, using the simplest possible discretization consistent with the topology of the surface while still getting the correct answer. This is convenient for our purposes, because 2D Yang-Mills theory on a genus-$g$ surface now becomes a simple lattice gauge theory with only $2g$ links (after pairwise identification), and only $2g$ link variables to integrate over [Fig.~\ref{fig:migdal}(b,c)]. In the topological limit $e^2=0$, the plaquette area $A_P$ does not enter the Boltzmann weight and only the topology of the discretization matters.

To compute the partition function on the genus-2 surface considered in the text, i.e., the volume $\vol(\BZ)$ of the $U(N$) Brillouin zone, we consider the discretization in Fig.~\ref{fig:migdal}(b) with a single octagonal plaquette:
\begin{align}\label{ZYMgenus2}
    \vol(\BZ)\equiv Z_\YM(0)=\prod_{\ell=1}^4\int dU_\ell\sum_Rd_R\chi_R(U_1U_2^{-1}U_3U_4^{-1}U_1^{-1}U_2U_3^{-1}U_4).
\end{align}
The Haar integrals can be performed by repeatedly using identity (\ref{ortho1}) as well as a second identity~\cite{BtD},
\begin{align}\label{ortho2}
    \int dU\,\chi_R(UVU^{-1}W)=\frac{1}{d_R}\chi_R(V)\chi_R(W),
\end{align}
and also the cyclic property of the trace (character). We thus obtain:
\begin{align}\label{EqVolBZ}
    \vol(\BZ)&=\sum_R\int dU_2dU_3dU_4\,\chi_R(U_2^{-1}U_3U_4^{-1})\chi_R(U_2U_3^{-1}U_4)\nn\\
    &=\sum_R d_R^{-1}\int dU_3dU_4\,\chi_R(U_3U_4^{-1}U_3^{-1}U_4)\nn\\
    &=\sum_R d_R^{-2}\int dU_4\,\chi_R(U_4^{-1})\chi_R(U_4)\nn\\
    &=\sum_R d_R^{-2},
\end{align}
where in the last line, we have used the orthonormality formula
\begin{align}\label{ortho3}
    \int dV\,\chi_{R'}(V^{-1})\chi_{R''}(V)=\delta_{R'R''},
\end{align}
which follows from Eq.~(\ref{ortho1}) upon setting $A=B=1$. More generally, for a genus-$g$ surface, we may use a discretization consisting of a single $4g$-sided polygon and edge identifications as in Fig.~\ref{fig:migdal}(c)~\cite{FarkasKra}. A similar exercise yields~\cite{witten1991,cordes1995}
\begin{align}\label{EqVolBZGenusg}
    \vol(\BZ)=\prod_{i=1}^g\int dU_idV_i\sum_Rd_R\chi_R([U_1,V_1][U_2,V_2]\cdots[U_g,V_g])=\sum_R d_R^{2-2g}.
\end{align}
The sum over $R$ on the right-hand side, known as the Witten zeta function, converges to a finite number for any $N\geq 2$~\cite{hasa2019} because the irrep dimensions $d_R$ grow rapidly and the exponent $2-2g$, which is the Euler characteristic of $\Sigma$, is negative for surfaces of genus $g\geq 2$. Finiteness of $\vol(\BZ)$ is expected from the compactness of $\BZ$~\cite{thaddeus1995}. As discussed in Appendix B of the main text, the Boltzmann weight in the topological limit (\ref{WPtop}) is in fact a delta function on the group manifold $U(N)$~\cite{BtD}:
\begin{align}
    \sum_Rd_R\chi_R(UV^{-1})=\delta(U,V).
\end{align}
In Eqs.~(\ref{ZYMgenus2}) and higher-genus analogs, this group delta function enforces the defining constraint $\c{R}(\{U_j\})=1$ of the $U(N)$ moduli space $\BZ$, where $\c{R}(\{\gamma_j\})$ is the relator appearing in the definition of the Fuchsian group $\Gamma$.

Although Migdal's lattice formulation is the one best suited to our work, one can also show the topological invariance of the 2D Yang-Mills partition function by working directly in the continuum~\cite{witten1991}. The Euclidean action functional is
\begin{align}
    S_\YM[A]&=\frac{1}{2e^2}\int_\Sigma\tr F\wedge\star F\nn\\
    &=\frac{1}{4e^2}\int_\Sigma\sqrt{\det\mathbb{g}}\tr F_{\mu\nu}F^{\mu\nu}\,\d x^1\wedge \d x^2\nn\\
    &=\frac{1}{4e^2}\int_\Sigma\omega\mathbb{g}^{\mu\lambda}\mathbb{g}^{\nu\rho}\tr F_{\mu\nu}F_{\lambda\rho},
\end{align}
where $\mathbb{g}_{\mu\nu}$ is a metric tensor on $\Sigma$ and $\omega=\sqrt{\det\mathbb{g}}\,\d x^1\wedge \d x^2$ is the associated volume form. Since $F_{\mu\nu}$ is antisymmetric, we can define a $\mathfrak{u}(N)$-valued 0-form $f$ via
\begin{align}
    F_{\mu\nu}=\sqrt{\det\mathbb{g}}\,\epsilon_{\mu\nu}f,
\end{align}
i.e., $F=f\omega$. Using the fact that the determinant of a $2\times 2$ matrix $M$ can be written as $\det M=\frac{1}{2!}\epsilon_{ij}\epsilon_{kl}M_{ik}M_{jl}$, we have
\begin{align}
    \mathbb{g}^{\mu\lambda}\mathbb{g}^{\nu\rho}\tr F_{\mu\nu}F_{\lambda\rho}=\epsilon_{\mu\nu}\epsilon_{\lambda\rho}\mathbb{g}^{\mu\lambda}\mathbb{g}^{\nu\rho}\det\mathbb{g}\tr f^2
    =2!\det\bigl(\mathbb{g}^{-1}\bigr)\det\mathbb{g}\tr f^2=2\tr f^2,
\end{align}
from which we obtain
\begin{align}
    S_\YM[A]=\frac{1}{2e^2}\int_\Sigma\omega\,\tr f^2.
\end{align}
Thus, the action functional does not depend on a choice of metric, only a choice of volume form $\omega$. The partition function $Z_\YM(e^2)$ is invariant under diffeomorphisms of $\Sigma$ that preserve the total area $A=\int_\Sigma\omega$.

Next, consider another $\mathfrak{u}(N)$-valued 0-form $\phi$, and the path integral
\begin{align}\label{pathintegral}
    \int\c{D}A\c{D}\phi\,e^{-\frac{e^2}{2}\int_\Sigma\omega\tr\phi^2-i\int_\Sigma\tr\phi F}
    &=\int\c{D}A\c{D}\phi\,e^{-\frac{e^2}{2}\int_\Sigma\omega\tr\phi^2-i\int_\Sigma\omega\tr\phi f}\nn\\
    &=\int\c{D}A\,e^{-\frac{1}{2e^2}\int_\Sigma\omega\tr f^2},
\end{align}
which is the Yang-Mills partition function $Z_\YM(e^2)=\int\c{D}A\,e^{-S_\YM[A]}$. Setting $e^2=0$ in the first line of Eq.~(\ref{pathintegral}), we find
\begin{align}
    Z_\YM(0)=\int\c{D}A\c{D}\phi\,e^{-i\int_\Sigma\tr\phi F},
\end{align}
which does not require the choice of a metric nor of a volume form on $\Sigma$. Thus, in the $e^2\rightarrow 0$ limit the theory is purely topological. This is manifest in the fact that the partition function in this limit, Eq.~(\ref{EqVolBZGenusg}), depends on $\Sigma$ only through its Euler characteristic $\chi(\Sigma)=2-2g$.

\subsection{Topological Wilson loops}\label{sec:WL}

Given a closed loop $C$ on $\Sigma$, we define the Wilson loop in a representation $R$ of $U(N)$ as the trace in this representation of the holonomy around $C$: $W_R(C)=\tr_R\bigl(\c{P}e^{i\oint_C A}\bigr)$. Here, we will be exclusively interested in Wilson loops in the fundamental ($F$) representation of $U(N)$, i.e., the $N$-dimensional defining representation. For simplicity, we denote such Wilson loops by $W(C)$:
\begin{align}
    W(C)=\tr_F\bigl(\c{P}e^{i\oint_C A}\bigr).
\end{align}
The Wilson loop is a gauge-invariant operator, thus its expectation value can be meaningfully computed as an insertion in the 2D Yang-Mills path integral:
\begin{align}
    \langle W(C)\rangle\equiv\frac{1}{Z_\YM}\int\c{D}A\,e^{-S_\YM[A]}W(C).
\end{align}
The computation of Wilson loop expectation values in 2D Yang-Mills theory is a rich subject in mathematical physics~\cite{witten1991,cordes1995,makeenko1979,kazakov1980,kazakov1981,rusakov1990,dahlqvist2022}. Here, we are only interested in Wilson loop expectation values in the topological ($e^2=0$) limit:
\begin{align}\label{WCexpval}
    \langle W(C)\rangle\equiv\frac{1}{Z_\YM(0)}\int\c{D}A\,\delta[F]W(C).
\end{align}
In this limit, simplifications occur, because then $\langle W(C)\rangle$ only depends on the homotopy class of $C$. That is why the identification $W(C)=\tr D^{(K)}(\gamma)$ in the main text is meaningful, although strictly speaking, it only holds for expectation values.

As discussed in Appendix C in the main text, to evaluate the functional integral (\ref{WCexpval}) in the Migdal formulation, we cut the Riemann surface $\Sigma$ along $C$ and introduce link variables along this curve~\cite{witten1991,cordes1995,rusakov1990}. The insertion $W(C)$ is then simply a character $\chi_F\bigl(\prod_{\ell\in C}U_\ell\bigr)$ in the fundamental representation, and its expectation value is
\begin{align}\label{WCexpvalMigdal}
    \langle W(C)\rangle=\frac{1}{Z_\YM(0)}\prod_\ell\int dU_\ell\,\chi_F\bigl(\textstyle\prod_{\ell\in C}U_\ell\bigr)\displaystyle\prod_P W_P.
\end{align}

To see that $\langle W(C)\rangle$ only depends on the homotopy class of $C$, consider the following example: a Wilson loop along the curve $C=U_1V^{-1}$ which encloses region $\alpha$ on the genus-$g$ surface $\Sigma$ in Fig.~\ref{fig:migdal}(c). Cutting the surface along this curve, we obtain two plaquettes $\alpha$ and $\beta$ such that $\Sigma=\alpha\cup\beta$. The corresponding Boltzmann weights are
\begin{align}
    W_\alpha&=\sum_\alpha d_\alpha\chi_\alpha(U_1V^{-1}),\\
    W_\beta&=\sum_\beta d_\beta\chi_\beta\Bigl(VV_1U_1^{-1}V_1^{-1}\prod_{i=2}^g[U_i,V_i]\Bigr),
\end{align}
where to avoid the proliferation of symbols, we reuse $\alpha$ and $\beta$ as irrep summation symbols for each plaquette. Equation~(\ref{WCexpvalMigdal}) then becomes
\begin{align}
    \langle W(C)\rangle=\frac{1}{Z_\YM(0)}\sum_{\alpha\beta}d_\alpha d_\beta
    \int dV\prod_{i=1}^g\int dU_idV_i\,
    \chi_F(U_1V^{-1})\chi_\alpha(U_1V^{-1})\chi_\beta\Bigl(VV_1U_1^{-1}V_1^{-1}\prod_{i=2}^g[U_i,V_i]\Bigr).
\end{align}
Using the fact that the character of a tensor product of irreps is the product of characters, and the decomposition of a tensor product into a direct sum of irreps~\cite{FultonHarris},
\begin{align}
    \chi_\lambda(U)\chi_\mu(U)=\chi_{\lambda\otimes\mu}(U)=\sum_{\rho\in\lambda\otimes\mu}\chi_\rho(U),
\end{align}
we have
\begin{align}\label{WCexample}
    \langle W(C)\rangle&=\frac{1}{Z_\YM(0)}\sum_{\alpha\beta}d_\alpha d_\beta\sum_{\rho\in F\otimes\alpha}
    \int dV\prod_{i=1}^g\int dU_idV_i\,
    \chi_\rho(U_1V^{-1})\chi_\beta\Bigl(VV_1U_1^{-1}V_1^{-1}\prod_{i=2}^g[U_i,V_i]\Bigr)\nn\\
    &=\frac{1}{Z_\YM(0)}\sum_\alpha d_\alpha\sum_{\rho\in F\otimes\alpha}\prod_{i=1}^g\int dU_idV_i\,
    \chi_\rho\Bigl(U_1V_1U_1^{-1}V_1^{-1}\prod_{i=2}^g[U_i,V_i]\Bigr)\nn\\
    &=\frac{1}{Z_\YM(0)}\sum_\alpha d_\alpha\sum_{\rho\in F\otimes\alpha}d_\rho^{1-2g}.
\end{align}
In the second line we have used Eq.~(\ref{ortho1}) to perform the integral over $V$, and in the third line, we have performed manipulations similar to those used in the derivation of Eq.~(\ref{EqVolBZ}) or Eq.~(\ref{EqVolBZGenusg}), i.e., repeated and alternated use of the character identities (\ref{ortho1}) and (\ref{ortho2}).

From Fig.~\ref{fig:migdal}(c), we see that the curve $C=U_1V^{-1}$ is contractible, since it is the perimeter of region $\alpha$ which can be shrunk to a point inside the polygon. Thus, in topological 2D Yang-Mills theory, we expect the expectation value $\langle W(C)\rangle$ to be equal to that for the trivial loop, $\langle W\rangle=\langle\tr_F(1)\rangle=\langle d_F\rangle=d_F$, where $d_F=N$ in the present $U(N)$ gauge theory. If this equivalence holds, we expect from Eq.~(\ref{WCexample}) that the following equality should hold:
\begin{align}\label{toprove}
    d_F\sum_\beta d_\beta^{2-2g}=\sum_\alpha d_\alpha\sum_{\rho\in F\otimes\alpha}d_\rho^{1-2g},
\end{align}
where we have used $Z_\YM(0)=\vol(\BZ)=\sum_\beta d_\beta^{2-2g}$ from Eq.~(\ref{EqVolBZGenusg}). To show that this is true, first define the tensor-product or Clebsch-Gordan coefficient $\c{N}_{\lambda\mu}^\rho$, which gives the multiplicity of irrep $\rho$ in the direct sum decomposition of the tensor product $\lambda\otimes\mu$:
\begin{align}\label{multiplicity}
    \chi_\lambda(U)\chi_\mu(U)=\chi_{\lambda\otimes\mu}(U)=\sum_{\rho\in\lambda\otimes\mu}\chi_\rho(U)=\sum_\rho\c{N}_{\lambda\mu}^\rho\chi_\rho(U).
\end{align}
Using Eq.~(\ref{ortho3}), this coefficient can be expressed as
\begin{align}
    \c{N}_{\lambda\mu}^\rho&=\int dU\,\chi_\lambda(U)\chi_\mu(U)\chi_\rho(U^{-1})\nn\\
    &=\int dU\,\chi_\lambda(U)\chi_{\rho^*}(U)\chi_{\mu^*}(U^{-1})\nn\\
    &=N_{\lambda\rho^*}^{\mu^*},
\end{align}
where for any representation $\rho$ we define the dual or conjugate representation $\rho^*$ such that $\c{D}^{(\rho^*)}(U)=\c{D}^{(\rho)}(U^{-1})^T$~\cite{FultonHarris}, from which $\chi_{\rho^*}(U)=\chi_\rho(U^{-1})$ and also $d_{\rho^*}=d_\rho$. Another result we will need is obtained by setting $U=1$ in Eq.~(\ref{multiplicity}):
\begin{align}
    d_\lambda d_\mu=\sum_\rho\c{N}_{\lambda\mu}^\rho d_\rho.
\end{align}
Using those results, the right-hand side of Eq.~(\ref{toprove}) becomes
\begin{align}
    \sum_\alpha d_\alpha\sum_{\rho\in F\otimes\alpha}d_\rho^{1-2g}&=\sum_{\alpha\rho}d_\alpha\c{N}_{F\alpha}^\rho d_\rho^{1-2g}\nn\\
    &=\sum_{\alpha\rho}d_\rho^{1-2g}\c{N}_{F\rho^*}^{\alpha^*}d_\alpha\nn\\
    &=\sum_{\alpha^*\rho^*}d_{\rho^*}^{1-2g}\c{N}_{F\rho^*}^{\alpha^*}d_{\alpha^*}\nn\\
    &=d_F\sum_\rho d_\rho^{2-2g},
\end{align}
which is the result (\ref{toprove}) we set out to prove. (We thank T.~Creutzig for suggesting this derivation.) Thus, we have shown (in this example) that $\langle W(C)\rangle$ is invariant under deformations of $C$ that preserve its homotopy class.

As discussed in the main text, the isomorphism $\Gamma\cong\pi_1(\Sigma)$ allows us to identify $W(C)=\tr D^{(K)}(\gamma)$. Since $Z_\YM(0)=\vol(\BZ)$, the Wilson loop expectation value (\ref{WCexpval}) is equal to the average of $\tr D^{(K)}(\gamma)$ over the non-Abelian BZ:
\begin{align}\label{WCexpval2}
    \langle W(C)\rangle=\int_\BZ\frac{dK}{\vol(\BZ)}\tr D^{(K)}(\gamma).
\end{align}
Thus, the $k$th moment (\ref{UNmoment}) of the $U(N)$ density of states can be expressed in terms of Wilson loop expectation values in topological 2D Yang-Mills theory:
\begin{align}\label{UNmomentWL}
    \rho_N^{(k)}=\frac{1}{N}\sum_{\alpha_1}\cdots\sum_{\alpha_k}\langle W(C_{\alpha_1}\cdots C_{\alpha_k})\rangle\tr T(\gamma_{\alpha_1})\cdots T(\gamma_{\alpha_k}).
\end{align}
In the main text, we consider the case $T(\gamma_\alpha)=1$ and thus $\tr T(\gamma_{\alpha_1})\cdots T(\gamma_{\alpha_k})=1$.

To compute $\langle W(C)\rangle$ with $C=C_{\alpha_1}\cdots C_{\alpha_k}$, we consider three different categories of loops, or more precisely their homotopy class $[C]\in\pi_1(\Sigma)$ which can also be thought of as a Fuchsian group element $\gamma\in\Gamma$.

\subsubsection{Contractible loops}

As discussed in the main text, contractible loops correspond to $\gamma=1$ where $1$ is the identity element of $\Gamma$. For such loops, Eq.~(\ref{WCexpval2}) directly gives
\begin{align}\label{contractibleWL}
    \langle W(C)\rangle=\int_\BZ\frac{dK}{\vol(\BZ)}\tr D^{(K)}(1)=N.
\end{align}

\subsubsection{Homologically nontrivial loops}\label{sec:homnontrivial}

Noncontractible loops correspond to $\gamma\neq 1$. Such loops can be further classified as homologically trivial if $\gamma\in[\Gamma,\Gamma]$, or homologically nontrivial if $\gamma\notin[\Gamma,\Gamma]$. Here $[\Gamma,\Gamma]$ is the commutator subgroup of $\Gamma$; recall that the first homology group of $\Gamma$ is the abelianization of $\Gamma\cong\pi_1(\Sigma)$, i.e., $H_1(\Sigma,\mathbb{Z})=\Gamma/[\Gamma,\Gamma]$~\cite{FarkasKra}. A homologically nontrivial loop (which is necessarily noncontractible) is such that when cut along it, $\Sigma$ does not separate into two different surfaces. In Appendix C in the main text, we use Fubini's theorem~\cite{BtD} to show explicitly that for such loops, $\langle W(C)\rangle$ vanishes for any $N$. It should be clear that the derivation therein holds not only for the $\{8,8\}$ lattice with octagonal fundamental polygon [Fig.~\ref{fig:migdal}(b)], but also for an arbitrary hyperbolic (Bravais) lattice with $4g$-sided fundamental polygon [e.g. Fig.~\ref{fig:migdal}(c)].

Another way to arrive at the same result is to invoke Weingarten calculus~\cite{weingarten1978,collins2003,collins2006,novaes2015}. Weingarten calculus allows one to calculate Haar integrals over compact Lie groups of polynomials in matrix elements $U_{kl}$, $U_{mn}^*$ of the fundamental (defining) representation of the group and its conjugate. The basic integral in the $U(N)$ case is
\begin{align}
    \c{I}_{p,q}\equiv\int_{U(N)}dU\,U_{k_1l_1}\cdots U_{k_pl_p}U_{m_1n_1}^*\cdots U_{m_qn_q}^*,
\end{align}
which depends on the four sets of indices $k_1,\ldots,k_p$, $l_1,\ldots,l_p$, $m_1,\ldots,m_q$, and $n_1,\ldots,n_q$. For $U(N)$, one has the ``neutrality condition''
\begin{align}\label{UNWg}
    \c{I}_{p,q}\propto\delta_{pq},
\end{align}
which essentially follows from integration over the center $Z(U(N))=U(1)$ as used in Appendix C. To show that this implies that $\langle W(C)\rangle$ vanishes, we first observe that arbitrary characters $\chi_R(U)$ of $U(N)$ can be expressed as poynomials in $\tr_F U^r$ and $(\tr_F U^r)^*$, where $r$ is a nonnegative integer~\cite{green1981,drouffe1983}. Thus, those characters are polynomials in the same matrix elements $U_{kl}$, $U_{mn}^*$ that appear in the Weingarten integral $\c{I}_{p,q}$. Now, since in $\c{R}(\{U_j\})$ each $U_j$ appears together with its inverse, $\chi_R\bigl(\c{R}(\{U_j\})\bigr)$ contains as many matrix elements of $U_j$ as matrix elements of $U^*_j$. However, for $\gamma_{\alpha_1}\cdots\gamma_{\alpha_k}\notin[\Gamma,\Gamma]$, the character $\chi_F(U_{\alpha_1}\cdots U_{\alpha_k})$ contains an imbalance of $U_j$ and $U_j^*$ matrix elements for at least one generator $j$. Thus, at least for this $j$, the Haar integral over $U_j$ will vanish because $p\neq q$ in Eq.~(\ref{UNWg}). This implies again that $\langle W(C)\rangle$ vanishes for homologically nontrivial loops.

\subsubsection{Homologically trivial noncontractible loops}

Homologically trivial noncontractible loops are those for which $\gamma\in[\Gamma,\Gamma]$ but $\gamma\neq 1$. Such a loop separates $\Sigma$ into two (punctured) surfaces when it is cut along it, i.e., it is the boundary of 2D submanifolds $\Sigma',\Sigma''\subset\Sigma$ such that $\Sigma=\Sigma'\cup\Sigma''$. In Appendix C, we leverage the mathematical results of Ref.~\cite{magee2021} to show that for such loops, $\langle W(C)\rangle\sim\mathcal{O}(1)$ in the large-$N$ limit.

\bibliographystyle{apsrev4-2}
\bibliography{main}